\documentclass[conference,compsoc]{IEEEtran}

\ifCLASSOPTIONcompsoc
  \usepackage[nocompress]{cite}
\else
  \usepackage{cite}
\fi
\ifCLASSINFOpdf
\usepackage{graphicx}
\else
\fi
\usepackage{url}

\usepackage[table]{xcolor}
\usepackage{booktabs}
\usepackage{hyperref}
\usepackage[most]{tcolorbox}

\newcommand{\revise}[1]{\textcolor{black}{{#1}}}

\newcommand{\change}[1]{\textcolor{black}{{#1}}}

\begin{document}
%
\title{When Privacy Becomes a Weapon: Understanding Doxxing and Privacy Vulnerabilities in \change{Mainland} China’s Social Media Ecosystem}

\author{
\IEEEauthorblockN{
Xiao Zhan$^{1,2,\dagger}$,
Shijing He$^{3,\dagger}$,
Chi Zhang$^{4}$,
Jose Such$^{5}$
}

\IEEEauthorblockA{
$^{1}$Universitat Politècnica de València, Valencia, Spain\\
$^{2}$University of Cambridge, Cambridge, United Kingdom\\
$^{3}$King's College London, London, United Kingdom\\
$^{4}$University of Warwick, Coventry, United Kingdom\\
$^{5}$INGENIO (CSIC-Universitat Politècnica de València), Valencia, Spain\\
$^{\dagger}$These authors contributed equally to this work.\\
Emails: xz603@cam.ac.uk, shijing.he@kcl.ac.uk, chi.zhang.9@warwick.ac.uk, jose.such@csic.es}
}


%


\maketitle

\begin{abstract}
Doxxing, the malicious disclosure of personal information, has become a pervasive privacy threat. Yet existing research remains predominantly Western-centric, limiting our understanding of how doxxing unfolds in contexts where mandatory identity systems, platform governance, and cultural logics fundamentally reshape privacy risks and harm trajectories. We address this gap through semi-structured interviews with 18 doxxing survivors in mainland China, synthesizing their experiences into a framework conceptualizing how doxxing operates in this context. %
Our findings reveal both 
\revise{patterns echoing prior Western findings,} such as platform amplification mechanisms that resonate with Western findings, and China-specific dynamics shaped by the interplay of regulatory mandates (compulsory identity linkage) and cultural logics including nationalist discourse, fandom culture, Confucian values, and low privacy literacy. Survivors' experiences further reveal how doxxing reshapes understanding of privacy: from preference to precondition, from momentary disclosure to temporal vulnerability, and from individual control to structural powerlessness. These insights challenge agency-centered privacy frameworks and suggest that effective protection requires constraining systemic vulnerabilities rather than relying solely on user empowerment. We conclude by proposing multifaceted recommendations spanning legal reform, platform design, and social initiatives. 
\end{abstract}


%
\IEEEpeerreviewmaketitle

\section{Introduction}
\label{sec:intro}

\revise{In contemporary social media ecosystems, personal information is continuously produced, stored, and searched across platforms. Doxxing exploits this condition by turning personal information into a resource for identification, reachability, and harm. Unlike data breaches that foreground organizational or technical compromise~\cite{ukgov}, doxxing is often socially driven: attackers draw on accessible traces, platform affordances, and interpersonal or collective conflict to expose or link information without consent. Once exposed, such information can circulate beyond the original attacker, enabling downstream harassment, stalking, reputational damage, and offline threats to survivors.\footnote{
We use "survivor" rather than "victim" for individuals who have been doxxed, following tech-abuse research~\cite{slupska2022networks,gupta2024really} and broader violence studies that emphasize diverse experiences of harm.}~\cite{zhou2024ethics,snyder2017fifteen,chen2019doxxing,douglas2016doxxing}}.

\revise{Empirical and conceptual work on doxxing has developed primarily in Western platform and legal contexts}
~\cite{douglas2016doxxing, chen2019doxxing, snyder2017fifteen}. This body of work has produced a foundation for understanding doxxing as a privacy attack, but its empirical base is geographically narrow. \revise{Comparative privacy research has shown that users in different national contexts hold systematically different concerns about online information disclosure~\cite{wang2011concerned,herbert2023world}, shaped by varying regulatory regimes, identity verification systems, and cultural norms. Whether and how these differences manifest in doxxing, how attacks are initiated, how survivors experience them, and what conditions enable them, remain largely unexamined.} 

\revise{Mainland China offers a useful case for extending this empirical base. Prior work has shown that China's digital ecosystem is shaped by mandatory real-name registration tied to national identity verification, established under the 2017 Cybersecurity Law and consolidated under the 2021 Personal Information Protection Law. Within this regulatory environment, research on Chinese social media users has noted self-censorship practices~\cite{chen2023we}, distrust of platform data handling~\cite{wang2016examining,zhou2024understanding,wang2015privacy}, and gaps between stated and actual privacy compliance~\cite{zhang2023understanding}. These features, particularly the real-name registration requirement, differentiate  Chinese from  Western social media ecosystems, yet whether or how they influence or facilitate doxxing  
remains unexamined.}

To address this gap, we conducted semi-structured interviews with 18 doxxing survivors in mainland China. We document survivors’ accounts of how doxxing unfolded, how it affected their understandings of privacy, and what roles platform infrastructures, regulatory arrangements, and cultural norms played in their experiences. 
Specifically, we address the following research questions:

    \textit{\textbf{RQ1} How do doxxing survivors experience and interpret doxxing incidents on Chinese social media platforms?}
 
    \textit{\textbf{RQ2} How does the experience of doxxing reshape survivors' understanding of privacy and vulnerability?}
    
    \textit{\textbf{RQ3} What role do Chinese culture and infrastructures play in doxxing?}
    

\textbf{Our contributions are:} 
(i) we document survivors' accounts of doxxing in mainland China, \revise{identifying findings both similar to and very different from prior Western findings, including the specific patterns due to the Chinese context; }
(ii) we demonstrate how doxxing transforms privacy from personal preference to matters of safety, structural vulnerability, and temporal exposure, calling for distinct theoretical frameworks; (iii) we identify how infrastructural mandates (real-name policy, mandatory IP display) and cultural logics (nationalist discourse, fandom culture, Confucian values, low privacy literacy) collectively enable and amplify doxxing in mainland China, \revise{arguing that their convergence constitutes what is distinctive about this context}; and (iv) \change{we develop an analytic framework connecting survivor-reported acquisition pathways, exposed information, and harms with infrastructural and sociocultural conditions, providing a structure for identifying intervention points and informing comparative research.}

\section{Background and Related Work}
\label{sec:related}


\subsection{Definition of Doxxing}
\label{sec:definition-doxxing}

The term ``doxxing'' traces its origins to 1990s hacker culture, derived from ``dropping documents/dox'', a retaliatory tactic involving the exposure of an opponent's offline identity~\cite{amada-def,beaujon-def}. \revise{Contemporary work generally conceptualizes doxxing as a privacy attack centred on the unauthorized or non-consensual disclosure of personal information by a third-party, rather than as generic online abuse~\cite{eckert2020doxxing,chen2019doxxing,karimi2022automated,douglas2016doxxing}. Douglas distinguishes three forms: deanonymizing doxxing links a pseudonymous identity to a real person; targeting doxxing reveals contact or location information; and delegitimizing doxxing exposes intimate or reputation-damaging information~\cite{douglas2016doxxing}. Across these forms, doxxing centers on the non-consensual disclosure, redistribution, or linkage of personal, sensitive, private, or identifying information, rather than hostility alone.}

\revise{Doxxing is often situated within broader discussions of cyberbullying and online harms~\cite{chen2019doxxing,zhou2024ethics,hinduja2014bullying}. However, whereas cyberbullying typically involves abusive interaction with a target, doxxing changes the target’s exposure conditions by making personal information available to audiences who can identify, contact, locate, shame, or threaten them~\cite{douglas2016doxxing,karimi2022automated}. 
Accordingly, in this study, we treat an incident as doxxing only when it involves non-consensual disclosure, redistribution, or linkage of personal identifying information; insults, rumors, threats, or manipulated media without such information exposure are treated as downstream or adjacent harms rather than doxxing itself.}


\subsection{Impact of Doxxing}
\label{sec:impact-of-doxxing}

Doxxing produces both first-order harms -- immediate harassment, reputational damage, physical threats -- and second-order harms that persist as chronic vulnerability and force survivors to police their digital behavior long after the initial incident~\cite{anderson2022harm}. The qualitative literature on survivors converges on a recurring pattern: beyond discrete harms, survivors describe pervasive uncertainty, loss of control, and the ineffectiveness of formal recourse through platforms or law enforcement~\cite{franz2023doxing,eckert2020doxxing}. These accounts also show that doxxing reshapes ongoing online participation -- through approach-avoidance coping, account withdrawal, or aggressive privacy adjustments~\cite{snyder2017fifteen,franz2023doxing} -- indicating that its impact is better understood as a sustained reconfiguration of digital life than as a bounded event. Parallel research situates doxxing alongside cyberbullying and stalking, where persistent digital hostility induces trauma-like symptoms such as hypervigilance, helplessness, and avoidance~\cite{chen2019doxxing,death-threat,douglas2016doxxing,rasheed2024influence}. Complementary work on perpetration links doxxing support to dark triad traits~\cite{darkdoxxing}. Across this body of work, however, the empirical base spans 
a limited range of contexts.

\subsection{Mitigation Strategies \& Legislation}
\label{sec:related-mitigation}

\noindent \textbf{Mitigation Strategies.} Doxxing-specific mitigation remains limited. 
\revise{Snyder et al.~\cite{snyder2017fifteen} built a pipeline to detect dox files on Pastebin, 4chan, and 8ch and characterize exposed information and subsequent harassment.} Karimi et al.~\cite{karimi2022automated} later evaluated automated approaches for detecting second- and third-party disclosures of sensitive private information on Twitter (now X). These studies focus on detecting exposed information, while adjacent interventions such as human--AI warnings~\cite{choi2023convex} or behavioral nudges~\cite{im2020synthesized} may reduce toxic escalation. However, they do not directly address doxxing’s core challenge: disclosed information can be copied, reposted, archived, and reused across platforms. To our knowledge, existing strategies have not been tailored to or evaluated in the context of Chinese social media.

\noindent \textbf{Legislation.} Legislative approaches to doxxing in Western jurisdictions remain predominantly fragmented, 
with most countries lacking a dedicated statutory framework.
The U.S. relies on state-level laws (e.g., in California and Washington), while the UK and Canada address doxxing indirectly through existing harassment and data protection laws. While Australia has empowered regulators via the Online Safety Act 2021, and recent moves in jurisdictions like the Netherlands and Australia are shifting towards specific criminalization, the prevailing Western model still requires a high threshold of establishing malicious intent or demonstrable harm. This reflects a persistent tension between privacy protection and freedom of expression.

\revise{Hong Kong has gone further with its 2021 Amendment to the Personal Data (Privacy) Ordinance, which directly criminalizes non-consensual disclosure of personal data~\cite{ong2025hong}. By contrast,} mainland China has not enacted legislation specifically targeting doxxing, and the practice is instead regulated implicitly through a patchwork of general privacy, public order, and data protection laws. Protections are distributed across the Civil Code, Criminal Law, Public Security Administration Law, and data-specific statutes such as the PIPL, Data Security Law, and the Law on the Protection of Minors. While this multi-layered framework provides a formal basis for regulating personal information misuse, its implicit and fragmented nature offers limited practical guidance for addressing doxxing as a distinct form of harm, leaving enforcement uneven and survivor protections uncertain. This paper considers and maps legal provisions that may apply to doxxing, distinguishing between regulations targeting doxxers and those governing platform responsibilities; these are summarized in Appendix \ref{appendix-legislation} to situate doxxing within existing legal frameworks.

\subsection{Privacy Risks in Chinese Social Media}
\label{sec:related-cyberbullyingCSM}

Privacy Research on Chinese social media has been limited, as both platforms and regulation have historically centred on state security and censorship. 
Within this context, personal data disclosure is often treated as an individual choice, and studies report relatively low stated concern about sharing. Wang and Yu~\cite{wang2016examining} and Zhou et al.~\cite{zhou2024understanding} found that many Chinese users are willing to share private data and appear relatively insensitive to privacy risks, even tolerating regulatory non-compliance~\cite{wang2016examining}. Yet behavior suggests greater caution: Chen et al.~\cite{chen2023we} found that 43\% of Weibo users self-censor to avoid privacy risks or cyberbullying, and Wang et al.~\cite{wang2015privacy} showed that users sometimes provide false information to firms and government platforms, indicating distrust not evident in survey responses.

Technical and institutional factors deepen this unease. Zhang et al.~\cite{zhang2023understanding} found that 15.65\% of data collection in 5,521 WeChat sub-apps were undisclosed, and only 34.4\% of data-collecting sub-APIs were covered by WeChat's permission system, 
while users remain sceptical of platform compliance with the PIPL due to continued exposure to opaque data practices~\cite{zhou2024understanding}.

\section{Methods}
\label{sec:method}

We conducted a qualitative study based on semi-structured interviews with 18 participants who had experienced doxxing 
in mainland China. 
All methods and procedures were approved by our Institutional Research Ethics Committee, and we discuss ethical considerations in detail in \S\ref{sec:ethics}.  
To promote open science, our study materials 
are publicly available in an \change{Zenodo} repository (\change{\url{https://doi.org/10.5281/zenodo.22832381}}).

\subsection{Recruitment}
\label{sec:recruitment}

Between July and August 2025, we shared recruitment calls across several social media platforms commonly used by Chinese users, including RedNote, WeChat Moments, and Douban. Interested individuals completed a Chinese-language screening survey covering demographics, doxxing incident details, and contact information for those willing to be interviewed. We recruited participants who:  1) were at least 18 years old; 2) had experienced doxxing on Chinese social media; and 3) resided in mainland China at the time of the incident. \revise{To ensure participants had actually experienced doxxing as 
defined in §\ref{sec:definition-doxxing}, we reviewed their reported incidents and excluded cases involving only insults, rumors, or harassment without disclosure of identifiable personal information.} From 120 screening responses, we sent interview invitations in small batches, prioritizing demographic balance and platform diversity. \revise{Recruitment stopped once no new codes emerged across four consecutive interviews, yielding a final sample of 18 (see \S\ref{sec:data-analysis}).} 


\subsection{Participants}

Table \ref{tab:participant-details} in Appendix presents detailed information about our interview participants ($n=18$). Our participants had a balanced gender distribution (9 male, 9 female) and ranged in age from 18 to 44. The majority ($n=11$) were aged 18–24, followed by $n=6$ aged 25–34, and one participant in the 35–44 age group. Educational backgrounds varied, with most holding or pursuing a Master's degree ($n=7$), followed by Bachelor's degrees ($n=6$), and others with an Associate Degree, High School education, or enrollment in university. Most participants were students ($n=12$), while others worked in fields such as technology, education, finance, freelancing, and design. Geographically, participants were distributed across a diverse range of provinces. 


\subsection{Interview Procedure}
We conducted all semi-structured interviews in August to October 2025, each denoted by a unique code $P\#$ (e.g., $P1, P2$). These interviews were conducted remotely via 
Microsoft Teams\footnote{\url{https://www.microsoft.com/en-us/microsoft-teams/}}. 
The duration of interviews ranged from 54 to 85 minutes (mean: 64.5 minutes). Participants were compensated at a rate of 150 yuan/h ($\approx \$20.9$/h). 
Each interview was led by one native Mandarin-speaking researcher, with a second researcher present in a non-leading role for note-taking. Only one researcher actively asked questions, and participants were informed of each researcher's role to minimize power imbalance. With informed consent of the participants, all interviews were audio-recorded and transcribed using Teams. 

During the interview, participants were first asked about their general social media usage, including the platforms they used, frequency of engagement, level of activity (e.g., posting, commenting), and identity presentation strategies (e.g., real-name vs. anonymous use). This initial section was designed as a warm-up to ease participants into the conversation, following best practices for interviewing on sensitive topics~\cite{muraglia2020conducting}. Participants were then invited to describe how they experienced and responded to the doxxing incident. They were asked whether they sought help from platforms or legal authorities, as well as the challenges they encountered and the outcomes of these efforts. Subsequent questions explored how participants' views and behaviors related to digital privacy had changed as a result of the incident, along with the broader impact on their offline lives. Finally, participants were asked to reflect on the societal and cultural factors within Chinese social media ecosystems that may contribute to or accelerate the occurrence of doxxing. The interview protocol was designed to be open-ended and participant-led. Participants were encouraged to speak freely and were invited to share any additional reflections, experiences, or advice that had not been addressed in earlier questions.

\textbf{Pilot Study.} We conducted two pilot interviews (not included in the final analysis) to gather feedback, which helped us refine the wording of our interview questions, particularly those related to participants' actual experiences of doxxing. 
Furthermore, pilot participants highlighted the impact of mandatory IP location displays. Therefore we added targeted questions to better explore this infrastructure, such as \textit{``Do the platforms you use mandatorily display your IP location?''}, and \textit{``What role do you believe this feature played in facilitating the spread of your personal information?''} 


\subsection{Data Analysis}
\label{sec:data-analysis}

We acknowledge sample size and determining data saturation remains a topic of debate in qualitative research~\cite{braun2006using,braun2021saturate}. In our study, we employed a continuous, iterative assessment process to determine this point: after each interview, the interviewers reflected on the data and met with the full research team to discuss emerging themes and assess data saturation~\cite{guest2006many}. 
We reached saturation after 14 interviews, when insights began to repeat and no new codes emerged, and conducted four additional interviews to confirm this. This final sample size aligns with widely accepted qualitative research standards, which prioritize rich, contextualized understanding over statistical generalizability~\cite{vasileiou2018characterising}.

We used reflexive thematic analysis~\cite{braun2019reflecting} to analyze the data from the semi-structured interviews. Prior to coding, all transcripts were read multiple times to develop a deep familiarity with the data. Two authors independently coded a randomly selected transcript and developed separate initial codebooks. They then met to discuss discrepancies, resolve disagreements, and merge their codebooks into a shared version. The two authors then independently applied it to two additional transcripts as a calibration exercise. Discrepancies in phrasing, scope, or abstraction were discussed to establish consistency in coding style, while allowing space for individual perspectives and reflexivity. Once aligned, the remaining transcripts were divided and coded independently, with openness to emergent codes. \revise{In weekly meetings throughout this phase, the two coders cross-reviewed each other's coded transcripts and emergent codes, disagreements were resolved through consensus, and any resulting code refinements were applied retrospectively across the dataset.} Theme development was further reviewed in broader team meetings involving the other two authors. 
We did not compute inter-rater reliability (IRR), as our analysis employed an interpretivist, consensus-driven approach in which the codebook was developed and refined through iterative discussions~\cite{ortloff2023different}. In line with McDonald et al.~\cite{mcdonald2019reliability}, we ensured analytic credibility through collaborative reconciliation meetings and systematic cross-checks of coding subsets, rather than relying on numerical agreement metrics.

In addition, the initial translation of the codebook and interview excerpts from Mandarin to English was carried out by the first author, a native Mandarin-speaking researcher. To ensure both linguistic accuracy and cultural fidelity, the second and third authors (also native Mandarin speakers) independently reviewed the translations to confirm that they were accurate and contextually appropriate~\cite{xian2008lost}.

\subsection{Author Positionality}




Our research team integrates interdisciplinary expertise across Computer Science, Political Science, and Law. Three authors are native Mandarin speakers born and raised in mainland China, with lived experiences in the Chinese socio-technical ecosystem and active engagement with the platforms discussed (e.g., Weibo, RedNote). This background enabled an insider perspective to decode implicit cultural norms (e.g., fandom, nationalism). Notably, the third author specializes in Chinese politics, focusing on security, nationalism, gender, and digital authoritarianism; and the fourth author, of European background, provided an outsider perspective that helped challenge cultural assumptions and ensure the accessibility for international audiences. Our team's gender diversity also enriched our interpretations of gendered harms. All authors are trained in Western academic institutions, positioning us at the intersection of Chinese cultural identity and Western theoretical frameworks. To mitigate potential bias, we held ongoing reflexive discussions to compare China-specific observations with broader literature, ensuring our interpretation remained grounded in participants' lived realities while being legally and theoretically rigorous.

\section{Findings}
\label{sec:findings}


\subsection{Survivors' doxxing Experiences (RQ1)}
\label{sec:context-of-doxxing}

\subsubsection{Attack Contexts}
\label{sec:attacktriggers}
\revise{
Most doxxing incidents began with online opinion disputes, including arguments around public events or nationalist topics. For instance, P4 was doxxed after commenting on a basketball game between China and another country, with attackers framing him as a \textit{``traitor (hanjian)''} before circulating his personal information across platforms (we analyze this dynamic in \S\ref{moral_framing}). Some doxxing incidents emerged from fandom conflicts, where criticism of a celebrity was framed as disloyalty to the fan community---as P3 recalled, \textit{``I just commented that I thought her historical costume looked better than her modern one, and he came at me, saying, `You think I can't find you?' and posted my personal info in the fan group.''} Doxxing incidents also arose from interpersonal or institutional relationships, such as breakups, workplace disputes, school conflicts, or transactional disagreements. In some cases, participants described doxxing as a form of moral punishment: attackers framed exposure as justified because the survivor had violated collective norms, damaged someone's reputation, or caused them to \textit{``lose face'' (mianzi)}---a culturally laden notion of social standing tied to family reputation that we \change{discuss further} in \S\ref{sec:culturalnorms}.}

\subsubsection{Doxxer Profiles}
\label{sec:doxxer-profiles}

\revise{Participants' accounts suggest two broad doxxer profiles. The first consisted of actors perceived as \textit{\textbf{technically capable, or able to access leaked or grey-market data}}. These actors were associated with more sensitive disclosures, such as addresses, ID numbers, or family information. For example, P9's ex-partner reportedly \textit{``had contacts who could, for a few thousand yuan, buy a person's entire profile, even relatives' details''}}.

\revise{The second, and more common, profile consisted of \textit{\textbf{ordinary users}}, including strangers in comment sections, fandom members, classmates, former partners, customers, or colleagues. These doxxers did not necessarily have privileged access or advanced technical skills. Instead, participants described them as relying on searchable traces, cross-platform profile matching, screenshots from acquaintances, visible metadata, institutional webpages, and purchasable personal-data services. P18 illustrated this pattern, describing how a doxxer with no special access pieced together his identity from username searches and old photos scattered across multiple platforms (we describe such pathways in \S\ref{sec:perceived-acquisition-pathways}). Thus, while China's real-name registration system does not mean that every doxxer directly accesses state or platform databases, it increases the linkability and value of leaked, semi-public, or contextually separated identity fragments once they enter circulation---a structural dynamic we examine in \S\ref{regulatory_mandates}.}

\revise{A few of participants further described the involvement of minors. We treat this not as a separate form of doxxing, but as evidence that some attacks were socially coordinated and low-barrier. P5, who eventually traced his doxxer's identity, was surprised to discover it was a secondary school student. Participants described minors as joining attacks for entertainment, group belonging, or a sense of justice---as P10 observed, \textit{``Minors might feel like they can just attack anyone they want, like it's no big deal, and they think that's something really cool.''} In a few cases, adults were perceived as encouraging teenagers to participate because they expected fewer consequences. P13 described a gaming streamer who routinely mobilized teenage followers: \textit{``[The streamer] guided or encouraged his followers to launch attacks or dox someone... mostly aged around 16 or 17, they are very easy to incite.''} These accounts show that doxxing in our data was not only an attack by highly skilled actors, but also a participatory practice that ordinary users---including minors---could join once identity cues had been surfaced.}

\subsubsection{Aggregated Exposure and Targetability}
\label{sec:aggregated-exposure}
\revise{Across all 18 participants, doxxing involved the non-consensual disclosure of personal identifying information. The most commonly exposed categories were real names, phone numbers, residential addresses, ID numbers, and school or workplace affiliations. Exposure was rarely confined to a single data point; doxxers typically circulated multiple categories at once,} so that a leaked phone number was accompanied by a real name, a workplace, or a profile photo. Some fragments had previously appeared in public or semi-public contexts, but participants experienced harm when these fragments were aggregated, removed from their original context, and redistributed to hostile audiences.

\revise{Exposure also extended beyond the survivor to their social ties. Several participants described doxxers disclosing information about parents, partners, or clients. }P13 recounted how a business rival exposed his parents' contact information, leading to sustained phone harassment: \textit{``People were calling and harassing us... especially my parents, they run a business, they couldn't ignore the calls because they never knew if it was a customer.''} In these cases, doxxing did not merely reveal who the survivor was, but mapped their social environment and created new channels for intimidation.

\revise{Participants further described the circulation of contextual and historical materials, such as screenshots of private chats, posts from semi-private groups, old social media content, school activity photos, or previously forgotten comments.} P8 reflected on how old posts were weaponized years later: \textit{``Those posts were from when I was much younger. Maybe immature, or just joking. But now they're taken seriously, out of context, by people who want to hurt me. It's like they're showing pages of a diary I never meant to share.''} The harm thus came not only from exposing hidden information, but from making past or context-bounded material newly searchable, interpretable, and actionable (we examine this temporal dimension further in \S\ref{sec:temporal}).

\revise{A few participants reported manipulated or synthetic content, including edited images or AI-generated sexualized material. In our data, such content became part of doxxing when attached to verified personal information such as a real name, account, school, workplace, or family relationship.} This combination made the fabricated material credible and damaging because audiences could connect it to an identifiable person. P13, for instance, reported that doxxers created pornographic videos by replacing his face with AI-generated deepfakes and posted them online alongside his real identity.

\subsubsection{Perceived Acquisition Pathways and Enabling Conditions}\label{sec:perceived-acquisition-pathways}
\revise{Participants described doxxing as the outcome of combining multiple acquisition pathways rather than any single disclosure. Individual fragments---a username, an old post, a contributor list, or a brokered phone number---could be innocuous or already public in their original context. Harm emerged when doxxers stitched these fragments together, anchored them to verified identity cues such as a real name, ID number, or residential address, and redistributed the assembled profile to hostile survivors. From participants' accounts, we identify four pathways feeding this assembly process. The first three were low-barrier and accessible to ordinary users (\S\ref{sec:doxxer-profiles}), while grey-market access was associated mainly with technically capable or institutionally connected actors. We briefly note the structural conditions participants saw as making each pathway effective, before returning to them in \S\ref{sec:social_media_and_cultural_norms_privacy}.}

\revise{\textbf{Cross-platform lookup.} Doxxers searched for usernames, profile photos, school names, fandom identities, or other recurring cues across multiple platforms, then linked the assembled fragments back to the original conflict space. Visible metadata such as IP-location tags and profile details narrowed the search rather than disclosing exact information on their own. P18 described this assembly process: \textit{``Even though I didn't post much on [the platform where the incident occurred], they used my username to track down some of my videos and old school activity photos from [another platform]. From those photos they figured out my school and my real name, then somehow got hold of my phone number---by the end they had pieced together a full profile of me, and started spreading rumors with all of that attached.''} Participants attributed the effectiveness of this pathway to two enabling conditions specific to their context: mandatory phone-number registration and real-name verification, which made identity fragments unusually linkable across platforms; and platform-mandated IP-location displays, which gave doxxers a credible geographic anchor without further investigation. We examine these regulatory and infrastructural conditions in \S\ref{regulatory_mandates}.}

\revise{\textbf{Mining historical archives.} Doxxers also retrieved older content---comments, photos, or posts that had disappeared from ordinary feeds but remained accessible through search, reposts, screenshots, or chronological browsing. Participants referred to the practical retrievability of public or formerly visible content. P9 contrasted platforms with different retention practices: \textit{``RedNote only shows recent posts, but Weibo is like an archive where you can dig up bad comments from many years ago.''} Participants did not experience these resurfaced posts as doxxing on their own; the harm came when old content was combined with newly disclosed contact or identity information, allowing hostile audiences to attach a past statement to a now-identifiable person. The enabling condition here was the absence of meaningful time-decay or searchability limits on legacy content, which we revisit in the context of platform design in \S\ref{risk_social_media}.}

\revise{\textbf{Acquaintance and institutional disclosure.} Some leaks originated from people or institutions in the survivor's environment. On the acquaintance side, ordinary users such as former partners, classmates, or colleagues forwarded screenshots from private or semi-private spaces, drawing on social proximity rather than technical capability. On the institutional side, universities, workplaces, public announcements, contributor lists, and activity pages exposed names, affiliations, awards, or addresses. P8 described how her university's public website listed students' names, awards, and activities, making her personal details \textit{``completely searchable''}, and P1 recalled how a magazine's contributor list had printed his full name and home address. These sources were rarely the sole basis of a doxxing incident; rather, they provided identity anchors that doxxers combined with other fragments. Participants linked the prevalence of institutional disclosures to the routine treatment of personal information as legitimately shareable in Chinese institutional and social settings (\S\ref{lack_awareness}).}

\revise{\textbf{Grey-market access.} Sensitive disclosures that participants could not attribute to any visible online source---phone numbers, residential addresses, family information, or ID-related data---were commonly attributed to grey-market data services. P4 noted that \textit{``there are sellers on Taobao who can get anyone's information through illegal means, some even claim to have friends in the police system who can trace IP addresses step by step.''} Participants understood back-end or institutional data as flowing through informal brokers and intermediaries, becoming accessible to ordinary attackers without requiring direct hacking. This pathway was primarily associated with the more technically capable or institutionally connected doxxer profile, though purchasable services lowered the barrier for some ordinary users as well. The scale and accessibility of this grey market reflect structural data-aggregation practices that we examine in \S\ref{risk_social_media}.
}

\subsubsection{Impacts and Consequences}
\label{sec:impact-consequences}


Beyond the privacy violation itself, which we analyze in depth in \S\ref{sec:changing-perceptions-of-privacy}. Participants reported a range of cascading harms that often co-occurred and reinforced one another. Emotional and psychological distress was the most prevalent consequence, with survivors describing fear, anxiety, insomnia, panic attacks, and persistent hypervigilance after their information circulated. For some, the distress was severe and prolonged; P6 recounted that being doxxed left him in a sustained state of vigilance that triggered \textit{``suicidal thoughts''} and required medication to manage stress-related heart symptoms.

Several participants experienced personal safety threats once contact or location information was exposed. P10 described receiving symbolic threats of physical harm: \textit{``They sent things like funeral wreaths or joss paper, even knives or other threatening items. Some might even show up in person, and in extreme cases, they could act on impulse, staging something like a car `accident' or other harmful incidents.''} Exposure also extended to family members, who became reachable to attackers and were drawn into the harassment (described in \S\ref{sec:aggregated-exposure}).

Social and reputational harm emerged when accurate identifying information was circulated alongside rumors, screenshots, or manipulated materials. P15 described how false narratives traveled quickly through her social and professional circles: \textit{``People who used to get along with me started looking at me strangely and avoiding me. The rumors spread quickly, even some of my clients heard it... a few projects were suspended, and even my mentor also suggested I take a break.''} Several participants further reported professional or academic disruption as a direct downstream effect, including suspended freelance work, advisor recommendations to step back from public visibility, or changes to their job or study environment. A smaller subset described tangible economic losses, such as lost income, costs of changing phone numbers or relocating, or expenses related to mental health treatment.

\subsubsection{Responses}
\label{sec:responses}

\textbf{\textit{Report incidents to social media.}} 
The most common response among participants was reporting incidents to social media platforms, in the hope that harmful posts would be removed or that the doxers' accounts would be suspended. Interestingly, a few participants described strategically leveraging social media censorship mechanisms (e.g., keyword filtering related to political or state security issues) when filing reports. They noted that standard categories like ``privacy breach'' or ``abuse/dispute'' were often addressed with significant delays or left unresolved altogether. As P11 stated: \textit{``I reported some doxers, but only one report was actually successful, and that was because the message was politically sensitive [...] I guess it was the phrase [a name of a Chinese national leader] that triggered the App's review system.''} 

\textbf{\textit{Limit legal knowledge \& support.}} 
Most participants had little to no awareness of the legal provisions that could be used for recourse
. Among the few who considered legal action, many doubted its practicality, as P15 noting that the absence of a dedicated anti-doxxing law makes it difficult to apply general provisions on personal information or public security. P10 further emphasized: \textit{``The cost of committing the crime is so low, but the cost of defending your rights is so high, like collecting evidence, or even figuring out who to turn to for help.
''} Moreover, several participants, including P14, specifically emphasized that the privacy of ordinary individuals is treated as cheap and expendable compared to that of wealthy or famous people. As P14 stated: \textit{``The laws and platforms always protect the celebrities first, not people like us. It's like our privacy doesn't matter as much, and honestly, justice just feels tilted toward the famous.''} This lack of a clear legal pathway reinforces the common perception that pursuing legal redress is neither feasible nor worthwhile.

\textbf{\textit{Retaliatory doxxing as a last resort.}} 
Some participants (7) attempted to retaliate by doxxing their perpetrators in return, describing these actions as a way to \textit{``fight back''} and reclaim agency in a situation that otherwise felt uncontrollable. While some framed these acts as a form of justice or deterrence, others saw them as the only available option in the absence of institutional protection. As P3 explained, \textit{``I felt like I had no other choice. I was forced to use the same tactics they used against me, because there was no effective system to protect me.''} \revise{Notably, none of these participants completed retaliatory doxxing; all stopped short of publishing information about the doxxer, citing limited practical value, legal risk, or doubts that retaliation would meaningfully deter the original attacker.} P18 reflected: \textit{``I did it in a moment of rage, but I stopped soon after. Looking back, it didn't improve my situation at all. I doubt the doxer even cared, someone who invades others' privacy probably doesn't care about their own.''}



\subsection{How doxxing Redefines Privacy and Vulnerability (RQ2)}
\label{sec:changing-perceptions-of-privacy}



\subsubsection{From Privacy as Choice to Privacy as Safety}
\label{sec:trauma-coping}

Before the doxxing incident, participants treated privacy as a matter of choice---managed through settings, platform decisions, and self-discipline, as P4 noted: \textit{``People who gets to see my information is up to me, no one else can find it.''}. This sense of agency translated into relatively mild precautions, such as reducing posting frequency or using protective tools. Privacy, at this stage, functioned as a controllable boundary between the self and others. Several participants further described this pre-doxxing period as one in which they could still enjoy social media as a space for self-expression and everyday documentation. As P9 reflected, \textit{``I didn't have many worries before, I would just post whatever I wanted to record my life''}. After the doxxing incident, however, P9 added: \textit{``Now I have to think and review multiple times before posting anything, speaking and even being a person no longer feels free''}.

\textbf{\textit{Post-doxxing visibility management.}}
After being doxed, participants no longer viewed privacy as a matter of discretion but as a means of survival. Their privacy practices became part of their perceived safety, with attempts to contain anxiety, and preempt exposure, and regain a sense of order in a world that had become unsafe. Most participants described either completely quitting the internet or temporarily avoiding all social media, even their mobile phones. For example, P8 shared that the sudden influx of strangers trying to add her after an incident led her to stop using the internet entirely. P6 further added: \textit{``I deleted everything on my other platforms [...] I thought better to just cut it off quickly---delete the comment before they even see it.''} A few participants further stressed that even seemingly neutral content, such as landscape photos or casual posts, became potential triggers. As P10 described avoiding posts entirely: \textit{``It's not that I mind people seeing my posts, but I just feel safer not sharing anything.''}

In contrast, a few participants proposed their practices: rather than quitting entirely, they strategically reduced visibility. For instance, P1 described his strategies facing doxxing: \textit{``First, I keep anonymous and make sure usernames are different across platforms, or using aliases and profile pictures, and also fill out as little personal info as possible. Second, I reduce connections between social media accounts. For example, Weibo lets you link to QQ or WeChat, but I avoid that.''} P16 further explained: \textit{``I post on WeChat Moments and interact with others much less often now, and I rarely appear in the photos myself''}, and she now carefully obscures or blurs identifying details (e.g., name) when sharing reels. 

\textbf{\textit{Privacy protection also need mental support.}} 
Many participants noted that after doxxing happened, they actively seek help to friends, family members, or professional psychological support. However, a few participants stressed that these supports were not always available or accessible, as P9 argued that the public institutions provide little mental or physical support for doxxing survivors. P8 added the sense of shame associated with seeking help from acquaintances or even family members, stating: \textit{``I'm someone who really cares about saving face, so I didn't want them to know about this [...] I was so furious and shame when I got back to my dorm, at that moment, I honestly felt like I didn't want to live anymore.''}


\subsubsection{Ghost from the Past}
\label{sec:temporal}

The doxxing experience also prompted participants to reflect on an aspect of privacy they had rarely considered before---its temporal nature. 
They described a dawning awareness that their digital past, once assumed to be irrelevant or forgotten, could suddenly be retrieved, recontextualized, and weaponized. 
As P9 reflected: \textit{``It felt like they were digging up my past just to shame me... posts I forgot even existed. They were exposing an old version of me I no longer recognize.''} Others explained that before being doxed, they had taken for granted that the internet would ``move on.'' Over time, new posts would bury the old, and outdated traces would simply lose relevance. doxxing shattered that assumption. Like P15 described: \textit{``Even if it was online before, that doesn't mean you can just drag it out years later. I didn't choose for it to live in front of me again---that feels like crossing a line.''} 

Participants also linked this temporal anxiety with specific platform affordances. For instance, some participants specially noted that platforms like Weibo with ``archival function'' uniquely dangerous: even without specialized databases, someone determined could still ``dig up your old dirt'' over a decade later. As P9 contrasted ``recent-only'' feeds with archives that make older content easily searchable: \textit{``RedNote only shows recent posts, but Weibo is like an archive where you can dig up bad comments from many years ago.''} 

For many, the act of resurfacing past content felt like a betrayal of an unspoken social contract---that publicness is contextual and temporary. P15, for example, recounted how a jealous colleague somehow assembled not only her private information but also details about her university ``dark past'' and then circulated this combination in spaces that generated a flood of sexualized harassment calls: \textit{``I have no idea where he posted my information to make me receive so many disgusting calls.''} While P8 further added: \textit{``Those posts were from when I was much younger. Maybe immature, or just joking. But now they're taken seriously, out of context, by people who want to hurt me. It's like they're showing pages of a diary I never meant to share.''} Through these accounts, participants reflected about privacy as something inherently bound to time. The doxxing incident made them aware that privacy is not only the ability to conceal information, but also the right to let the past remain in the past. 

\subsubsection{From Personal Boundaries to Asymmetric Technical Power Relations} 
\label{sec:boundary-exposure}

Rather than experiencing doxxing merely as a violation of personal boundaries, participants described it as an encounter with asymmetric technical power relations, in which attackers could uncover, assemble, and weaponize personal information in ways that individuals could not realistically defend against. In this context, privacy was no longer defined by what one chose to disclose, but by what others had the technical capacity and willingness to extract and mobilize.

Doxxing was experienced as a targeted process of producing visibility, rather than an incidental outcome of everyday data sharing. Attackers actively searched across platforms, mined historical content, and recombined fragments of information to construct harmful narratives. As P9 explained: \textit{``The internet has no memory' is a lie [...] Privacy isn't about what I choose to share any more, it's about what others can dig up, piece together, and use against me.''}
This distinguishes doxxing from interdependent privacy loss, where exposure is often unintended or diffuse. This asymmetry was further amplified by attackers’ access to tools and infrastructures unavailable to ordinary users. Participants described doxxers as operating anonymously, at scale, and with minimal risk, while survivors faced persistent exposure without effective means of defense. P5 noted, \textit{``They don’t even need to know me. They just run a bit of code and pull up all your info.''} The growing availability of grey-market data services and automated tools was seen as lowering the barrier to doxxing, transforming it into a routine and repeatable practice.


Emerging technologies intensified this imbalance. Participants expressed concern that generative AI and search-based summarization tools made it easier to infer identities, reconstruct profiles, or generate convincing but fabricated content. As P15 observed, \textit{``Before, you had to dig, maybe learn some coding. Now you just type a few words, and it tells you everything---even fake stuff. Anyone can do it now.''} A few participants, like P9 described AI summarization features in some social media platforms that recombine their past posts into an \textit{``inferred profile''}. 
These experiences reframed privacy from a matter of personal control to a condition shaped by structural and technological power beyond the individual, rendering individual defenses largely ineffective.

\subsection{The Role of Chinese Infrastructure and Culture in Facilitating doxxing (RQ3)} \label{sec:social_media_and_cultural_norms_privacy}

\subsubsection{Regulatory Mandates Enable Doxxing}
\label{regulatory_mandates}

\revise{Participants identified two state-mandated regulatory features that they perceived as structurally enabling doxxing: real-name registration and mandatory IP-location display on social media platforms.}

\textbf{\textit{Real-name registration and cross-platform binding.}} China's real-name registration system requires users to verify their identity---typically through a mobile number tied to a national ID---before accessing all major platforms. \revise{Participants described this as creating a single, traceable identity that linked their activity across services, so that any leaked fragment could be anchored to a verified legal identity.} As P5 explained: \textit{``If a doxxer hacks one platform and gets your data, they can use the same credentials to log in everywhere else.'' }This binding made the cross-platform stitching described in \S\ref{sec:aggregated-exposure} substantially more effective than it would be in pseudonymous environments: a username, phone number, or school affiliation surfaced on one platform did not stay confined to that platform but could be re-anchored to the same legal person across the wider ecosystem. Most participants noted that reducing such linkage was ``technically difficult'' and ``rarely attempted'', leaving them effectively bound to a unified digital identity.

\textbf{\textit{Mandatory IP-location display.}} Platforms such as Weibo and RedNote automatically display each user's IP-based geographic location alongside posts, with no option to disable the feature. \revise{Although this display reveals only a provincial or city-level location rather than a precise address, participants described it as providing doxxers with a credible geographic anchor that narrowed downstream investigation. }P17 noted: \textit{``I don't see the benefit of showing my IP location, and some platforms take it to the extreme. Even if you're not posting, just logging in shows your IP. And with that, a doxxer can easily figure out where you live.''} \revise{Combined with profile details, posting histories, or school affiliations, this regulator-mandated metadata removed a step doxxers would otherwise have to take through investigation or paid services (discussed further in \S\ref{sec:mandatory-identity-infras}).}


\subsubsection{Risks from Social Media Platforms} \label{risk_social_media}
Beyond regulatory mandates, participants identified two categories of platform-level conditions that enabled doxxing. 

\textbf{\textit{Data access through extraction and aggregation.}} 
Participants described platforms as engaging in continuous background data harvesting that built persistent cross-application profiles---datasets that could be exploited for doxxing once leaked or sold. P1 and P2 reported that RedNote attempted to access their device clipboard\footnote{The device clipboard is a temporary system buffer that holds recently copied content across applications. Repeated access allows an app to observe data the user has handled but never explicitly shared with it.} \textit{``hundreds of times per day''}, a behavior they interpreted as covert surveillance. P1 further observed that this tracking appeared to extend beyond a single application: \textit{``I was shocked that RedNote could even pull data from other apps. I booked a flight elsewhere, and suddenly RedNote was showing me travel guides for that city. It feels like RedNote secretly extracts and stores data from my phone to build a personalized knowledge base about me---information I never disclosed on the platform itself. That gives any doxxer an even richer map of my private life to exploit.''} Participants linked this extraction to the broader infrastructure that fed the grey-market access pathway (\S\ref{sec:perceived-acquisition-pathways}). As P5 argued: \textit{``They connect all the scattered personal information we report online, which produces one social-engineering database after another, specialized in doxxing.''} In this view, the harm was not produced by individual hackers but by formal data collection practices whose outputs became accessible to ordinary attackers through informal markets.

\textbf{\textit{Selective enforcement and unaccountable amplification.}} Even when primary doxxers were occasionally sanctioned, participants described platform enforcement as systematically failing to address the broader networks of users who reposted, amplified, or extended attacks after initial exposure. \revise{These secondary participants spread rumors, posted insults, or disclosed additional personal information, but rarely faced consequences and often disappeared into anonymity once incidents subsided. P7 captured this asymmetry: \textit{``If you are not the first one, you can get away with it.''} Several participants described this selective enforcement as signaling that opportunistic harassment carried little risk, encouraging continued participation in cascades that survivors could not contain.} 

\subsubsection{Moral Framing and Social Acceptance of Doxxing} \label{moral_framing}

Many participants described that doxxing incidents were often initiated or escalated by a central actor or small group, who deliberately framed the event in moral or ideological terms---as a matter of justice, loyalty, or social correction. This moral reframing, they explained, created a context in which privacy violations became rhetorically justified. For instance, P12 explained that once such framings took hold, comment sections and group threads quickly formed one-sided moral consensus, leaving little space for dissent, stating: \textit{``When most people in the comments agree it's justice, no one dares to say it's wrong.''} 
\revise{Two moral frames recurred across participants' accounts: nationalist discourse and fandom loyalty.}

\textbf{\textit{National framing.}} 
Many participants situated doxxing within nationalist moral frames, where disagreement was quickly recast as a failure of patriotism and thus as grounds for punishment. They reported that contesting one's exposure often backfired: any attempt to reassert personal boundaries was read as arrogance, defiance, or further proof of ``guilt,'' prompting attackers to double down under the banner of justice for the nation. For example, P4 described being doxxed by self-described ``patriotic youth'' after arguing about a basketball game between China and another country (which China lost): \textit{``I just posted some comments about the game, and people with opposing views started arguing under my post. They accused me of being unpatriotic, even called me a `traitor (hanjian)'. It then turned into personal attacks on me and my family […] soon they were posting my information on livestream apps and other platforms.''} As P18 further explained: \textit{``If they [the doxxers] think you deserve it, fighting back only makes them more sure.''} Through these dynamics, participants came to see nationalist discourse not as a backdrop but as a key trigger and justification for doxxing, turning exposure into a form of collective punishment in which privacy claims could be dismissed as anti-national or morally suspect.

\textbf{\textit{Fandom culture.}} 
Participants specifically highlighted fandom culture as the key context in which moral justifications were frequently mobilized to legitimize doxxing. In fan communities, even mild criticism of a celebrity could be construed as an attack, triggering retaliatory behavior framed as \textit{``defending the idol.''} \revise{Within these tightly organized groups, doxxing became a way of demonstrating loyalty and enforcing boundaries against perceived disrespect.} P11 specifically noted that many fandom-driven doxxers were under 18 and predominantly female, often engaging in doxxing behavior against those they perceived as showing \textit{``disrespect''} toward their idols. P3 further recalled: \textit{``I just commented that I thought her historical costume looked better than her modern one, and he came at me, saying, `You think I can't find you?' and posted my personal info in the fan group.''} \revise{In this configuration, \textit{``defending the idol''} both justified participation and rewarded it socially, turning fandom doxxing into a coordinated performance of group loyalty rather than the act of a single skilled doxxer.}

\begin{figure*}
    \centering
    \includegraphics[width=\linewidth]{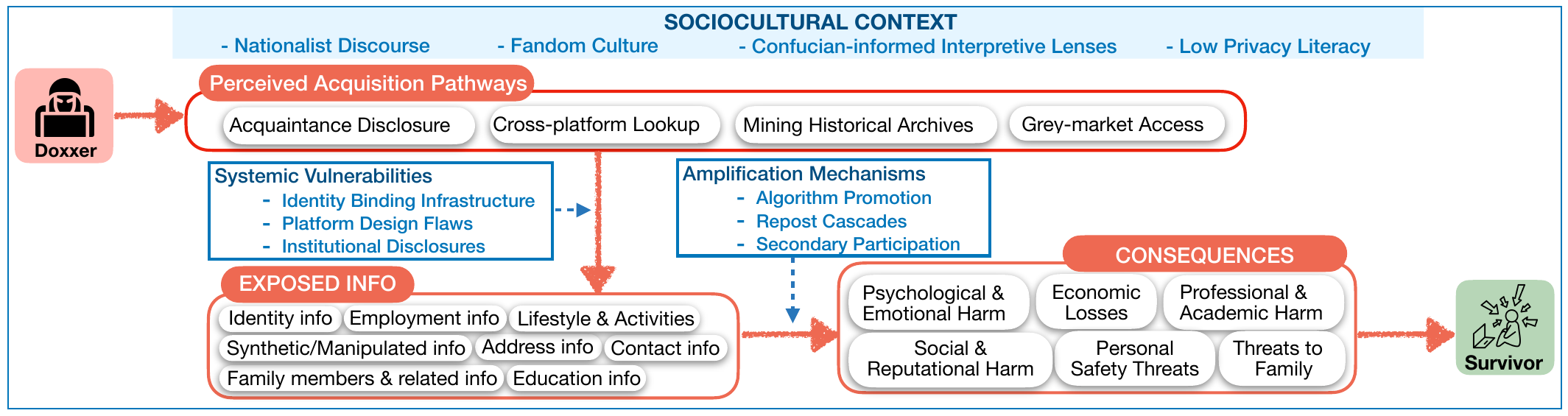}
    \caption{\change{An analytic framework of doxxing grounded in survivors’ accounts in mainland China. Red arrows trace the core process, and blue dashed arrows link systemic vulnerabilities and amplification mechanisms to process transitions. The enclosing frame represents the sociocultural context.}}
    \label{fig:threat-model}
\end{figure*}

\subsubsection{Lack of Privacy Education \& Cultural Awareness} 
\label{lack_awareness}

Many participants attributed the normalization of doxxing in Chinese digital spaces to a broader lack of privacy education and weak cultural awareness of digital boundaries \revise{-- conditions that, in their view, made it possible for ordinary users to participate in or condone doxxing without recognizing it as harm.}

\revise{Participants described formal education as offering little preparation for managing personal information online.} As P1 recalled, \textit{``Our education never really covered how to protect this kind of information, it was just a blank space.''} P5 and P6 similarly criticized the idealistic tone of school education, which emphasized societal trust and goodness without preparing individuals to recognise harm: \textit{``We were taught that the world is full of kindness and beauty. No one ever told us there's danger out there, [...], we just fell into traps without knowing.''} 

\revise{Beyond formal education, }participants pointed to a wider cultural environment that normalizes the downplaying of privacy. P9 emphasized that \revise{intrusive questioning is treated as ordinary within familial and social networks:}\textit{``People think it's normal to ask about your relationship status, your job, even private stuff. Kids aren't seen as needing privacy, parents think they have the right to know everything.''}\revise{P16 traced the same conditioning to institutional practices that began in childhood, }\textit{``Since we were little, we had to fill out all kinds of forms---school, neighborhood, everything---sharing our personal details without being asked if we wanted to. Eventually, it just became automatic. Even now, when a website asks for private information, we fill it in without thinking twice.''} \revise{This habitual overexposure shaped not only how participants treated their own information, but also how bystanders interpreted others' information. P15 reflected on the resulting complicity: \textit{``They don't think they're doing anything wrong. They just think, 'I'm sharing someone's info, so what?' They don't realise that makes them complicit.''}}


Several participants further connected this normalization to the cultural framing of privacy (yinsi) itself. \revise{In everyday Chinese discourse, yinsi often carries a negative connotation, implying selfishness, pettiness, or something improperly concealed.} 
P1 reflected, \textit{``This phenomenon may be related to China's thousands of years of traditional culture [...] When I was in medical school, I often saw people who liked to pry into others' privacy, like what illness someone had.''} \revise{In this configuration, asserting privacy becomes socially costly, and complaints about doxxing risk being dismissed as oversensitivity---a dynamic that not only discouraged survivors from seeking support (\S\ref{sec:trauma-coping}) but also lowered the social cost of doxxing for the ordinary users who participated in it.}

\section{Discussion}
\label{sec:discussion-new}


\subsection{Conceptualising Doxxing}
\label{sec:conceptualising-doxxing}

\change{Grounded in participants’ accounts, we propose an analytic framework (Figure \ref{fig:threat-model}) that connects information acquisition, exposure, and harm within mainland China's socio-technical context. Red arrows trace the core process from the doxxer to survivor through three components shown in red-outlined rounded boxes: perceived acquisition pathways, exposed information, and consequences. Perceived information acquisition pathways capture how participants reported or inferred that their information was obtained (\S\ref{sec:perceived-acquisition-pathways}). Exposed information describes the identifying and contextual material circulated (\S\ref{sec:aggregated-exposure}). Consequences capture the overlapping harms survivors experienced (\S\ref{sec:impact-consequences}). Multiple acquisition pathways can contribute to an identifiable profile whose circulation enables harassment and other harms.}

\change{The enclosing blue frame situates the entire process within its sociocultural context, while the two blue rectangular boxes identify systemic vulnerabilities and amplification mechanisms. Blue dashed arrows connect these conditions to the transitions from acquisition to exposure and from exposure to harm, respectively. Systemic vulnerabilities—including identity-binding infrastructure, platform design flaws, and institutional disclosure practices (\S\ref{regulatory_mandates}, \S\ref{risk_social_media}, \S\ref{sec:perceived-acquisition-pathways})---concern what information is accessible and how readily fragments can be linked to individuals. Amplification mechanisms—including algorithmic promotion, repost cascades, and secondary participation (§\ref{risk_social_media})---concern how exposed information reaches wider audiences and becomes a basis for continued harassment. Sociocultural context encompasses nationalist discourse, fandom practices, and limited privacy literacy (\S\ref{moral_framing}, \S\ref{lack_awareness}). We also draw on mianzi and filial piety as possible interpretive lenses for participants’ accounts of reputational harm, help-seeking, and family-directed exposure (\S\ref{sec:culturalnorms}).}

\change{The framework provides a structure for tracing doxxing incidents and identifying intervention points (\S\ref{RECOMMENDATION}). For example, reducing the visibility of identity-linked information addresses acquisition, while limiting repost cascades addresses the spread of exposed information. It also offers a starting point for comparative research. The following sections distinguish patterns shared with prior Western research (\S\ref{norms_China_Western}) from features salient in our mainland Chinese setting (\S\ref{doxxing_China_unique}). Studies in other contexts can use these components to examine which pathways and harms recur, while adapting the contextual dimensions to local identity arrangements, platform practices, and social norms.}

\subsection{Patterns Shared with the Western Context} \label{norms_China_Western}

Using the framework introduced above, we observe that two patterns emerging from participants' accounts align with prior work on doxxing in Western contexts.

\revise{\textbf{\textit{Consequences and survivor responses}} echo dominant findings in the Western survivor literature. Participants' accounts of uncertainty, loss of control, hypervigilance, ineffective recourse, and wholesale withdrawal from platforms (§\ref{sec:impact-consequences},§\ref{sec:responses},§\ref{sec:trauma-coping}) align with patterns documented in interview-based~\cite{eckert2020doxxing} and large-scale quantitative studies~\cite{snyder2017fifteen} of Western survivors.}

\revise{\textbf{\textit{Amplification mechanisms}} are also shared: attention-based business models amplify high-engagement content regardless of context~\cite{milli2025engagement,huszar2022algorithmic,kumar2023understanding}, and primary doxxers occasionally face sanctions while secondary participants who repost or extend attacks remain unaccountable -- a dynamic documented on Western platforms~\cite{kim2024respect,thomas2022s,kim2022does} and present in our data (§\ref{risk_social_media}).}

Though not intended by design, these mechanisms become systematic enablers when weaponised for doxxing. 

\subsection{Conditions \change{in the Mainland Chinese Context}}
\label{doxxing_China_unique}

\change{We now discuss two contextual dimensions reflected in participants’ accounts: mandatory identity infrastructure (\S\ref{sec:mandatory-identity-infras}) and the sociocultural context of exposure and harm (\S\ref{sec:culturalnorms}).}


\subsubsection{Mandatory Identity Infrastructure as Structural Enabler} 
\label{sec:mandatory-identity-infras}

\revise{The mandatory identity requirements documented in \S\ref{regulatory_mandates}---real-name registration tied to national ID, and forcibly displayed IP-location metadata---function not as ordinary platform-level vulnerabilities but as \textbf{\textit{systemic preconditions}} of online participation. We characterize the resulting condition as \textbf{\textit{compulsory legibility}}: a state in which users cannot opt out of being identifiable without exiting the digital ecosystem entirely. This shifts the analytical question for doxxing in this context. Where prior conceptual work frames doxxing primarily as the revelation of a previously hidden link between online identity and offline person~\cite{douglas2016doxxing}, compulsory legibility removes that boundary at the regulatory level: because online and offline identities are already linked, doxxing exploits this prior linkage rather than constructing it. This reframing has two analytical consequences. First, the attack model for doxxing changes. In pseudonymous environments, doxxers must invest effort in \textit{establishing} the link between online identity and legal person; in compulsorily legible environments, this link is already maintained by the infrastructure, and the doxxer's task reduces to surfacing information that is already implicitly connected to a verified identity. This helps explain a pattern recurring across participants' accounts: many doxxers were ordinary users without technical capability (\S\ref{sec:doxxer-profiles}), yet were able to construct full identity profiles by combining fragments across platforms (\S\ref{sec:perceived-acquisition-pathways}). That is, compulsory legibility \textbf{lowers the technical threshold} of attack and makes it much easier in general.}

Second, the regulatory context produces an \textbf{\textit{asymmetric protection regime}}. The same infrastructure that mandates identifiability for state oversight provides no corresponding protection against peer-to-peer weaponization of that identifiability. Participants experienced this as a paradox (\S\ref{sec:responses}): authorities treat doxxing as an ``online dispute'' unworthy of intervention, while the structural conditions that make doxxing easy are themselves state-mandated. This contrasts with regulatory developments in jurisdictions such as Hong Kong~\cite{ong2025hong} and Australia~\cite{Australia}, where doxxing is increasingly recognized as actionable harm. In contrast, in mainland China, doxxing remains regulated implicitly through general privacy and public order statutes (Appendix \ref{appendix-legislation}), leaving survivor recourse uncertain.

\subsubsection{Cultural Logics Legitimize \& Intensify Doxxing}
\label{sec:culturalnorms}


\change{Beyond the infrastructure, we consider how sociocultural context may help explain the justifications for doxxing and the meanings of exposure and harm in survivors' accounts.}

\textbf{Nationalist framing as collective punishment.} The nationalist double bind we documented (\S\ref{moral_framing})---where contesting accusations of disloyalty intensifies them---is not merely a rhetorical pattern but reflects a structural alignment between state discourse and platform governance~\cite{jiang2018chinese}. When national security consistently overrides individual privacy protections in both regulatory frameworks~\cite{he2025privacy,calzada2022citizens,creemers2022china} and platform moderation, nationalist framing acquire structural legitimacy that other moral framings lack. These shifts doxxing from individual privacy violation into a form of collective punishment that can be weaponized against ordinary users over mundane disputes. \revise{Rather than appearing as a common or recurring mechanism, this dynamic is less frequently documented and more related to public figures in the existing Western literature, e.g., 
police officers being doxxed in Canada~\cite{huey2025cops}.}

\textbf{Fandom culture.} What participants experienced as ``quasi-military'' fandom operations (\S\ref{moral_framing}) reflects a configuration in which platform economic incentives and cultural mobilization mutually reinforce. Hierarchical fan groups, coordinated metrics, and monetized engagement~\cite{dan2023observations,zhai2023fans} turn devotion into infrastructure for collective action, including doxxing as boundary enforcement against perceived disrespect. \revise{While Western literature has documented fan-driven networked harassment (e.g., review bombing, coordinated trolling, mass reporting)~\cite{marwick2021morally,stanfill2024fandom}, doxxing as a coordinated performance of fandom loyalty appears largely specific to the Chinese ``fan quan'' ecosystem, where formalized commercial and organizational structures channel collective action into sustained identification campaigns.}


\change{\textbf{Confucian values as a possible lens on reputational harm and help-seeking.} The Confucian concept of \textit{mianzi (face)} may help interpret two related patterns in participants' accounts: reputational harm and reluctance to seek support (\S\ref{sec:trauma-coping}, \S\ref{lack_awareness}). Exposure can threaten both individuals' social standing and their family's reputation, concerns discussed in prior studies on face and shame~\cite{bedford2003guilt,li2004organisation}. Concerns about face may also make seeking support more difficult: P8 described not wanting acquaintances of family members to know about the incident because of the importance of \textit{``saving face''} (\S\ref{sec:trauma-coping}).}

\textbf{Confucian values as a bidirectional harm vector.} The Confucian concept of \textit{mianzi (face)} operates not as a uniform cultural backdrop but as a \textit{bidirectional dynamic} that simultaneously amplifies harm and forecloses recourse (\S\ref{sec:trauma-coping}, \S\ref{lack_awareness}). On one side, doxxers weaponize the threat of social disgrace, threatening not only the individual but their family's standing~\cite{bedford2003guilt,li2004organisation}. On the other hand, the same concern for face deters survivors from seeking help, since acknowledging exposure means acknowledging loss of face.

\change{\textit{Filial piety and kinship-based vulnerability.}
Survivors described how relatives' information could be obtained (P9) and how parents were subjected to harassment (P13) (\S\ref{sec:aggregated-exposure}), suggesting that family ties could themselves become a source of vulnerability. Filial piety offers a possible lens for understanding the significance of such exposure: perceived obligations toward parents and responsibility for family honor~\cite{chan2024confucian} may give threats against relatives additional emotional and moral weight. Viewed alongside the identity infrastructure discussed in \S\ref{sec:mandatory-identity-infras}, these accounts suggest how relational and infrastructural vulnerabilities may intersect: identity linkages can make relatives reachable, while family obligations may heighten the significance of their exposure for survivors.}

\textit{Filial piety as kinship weaponization.} A related but distinct dynamic emerged in participants' accounts of deliberate targeting of parents and family members (\S\ref{sec:aggregated-exposure}). Where Western survivor literature~\cite{eckert2020doxxing,franz2023doxing} has primarily documented harm radiating outward from the survivor as secondary effect on social ties, doxxers in our data strategically targeted family members as primary points of attack, exploiting the Confucian framework in which children bear moral responsibility for family honor~\cite{chan2024confucian}. This weaponization is structurally enabled by the broader identity infrastructure (\S\ref{sec:mandatory-identity-infras}) that makes family relationships linkable through real-name records and shared phone numbers, turning kinship into an attack vector rather than collateral exposure.

\textbf{Low privacy literacy.}
The institutional and cultural normalization of personal information sharing (\S\ref{lack_awareness}) functions less as an independent dynamic than as a \textit{substrate} that enables the other three. \textbf{\textit{Institutional practices that treat disclosure of personal information as routine intersect with cultural norms that treat such information as legitimate conversational material}}, leaving the boundary between ``sharing'' and ``exposing'' weakly demarcated. As a result, the secondary and bystander participation that amplifies attacks (\S\ref{risk_social_media}) appears unremarkable to those involved~\cite{he2025exploring}, lowering both the social cost of joining a doxxing cascade and the social resistance survivors can mobilize against it.


\subsection{Rethinking Privacy through Doxxing}

Survivors' reflections (§\ref{sec:changing-perceptions-of-privacy}) reveal that when personal information is weaponized, privacy operates through dimensions existing frameworks do not fully capture
. We identify three shifts that extend current theoretical understandings. 

First, privacy shifted from preference to precondition. Privacy is treated as one value individuals balance against convenience, sociability, or transparency, assuming meaningful choice over disclosure~\cite{westin1968privacy,trepte2014people,nissenbaum2004privacy}. These frameworks work when privacy is negotiable. But survivors described privacy becoming existential: participants who previously navigated disclosure trade-offs adopted wholesale withdrawal because privacy no longer meant managing visibility but ensuring survival (\S\ref{sec:trauma-coping}). This reveals privacy's hierarchical nature: when weaponized, it transforms from one consideration among many into a prerequisite for participation itself.

Second, privacy became understood as temporal. Survivors realized vulnerability stemmed not from what they shared but from what could be retrieved and recontextualized years later (\S\ref{sec:temporal}). Existing frameworks treat privacy as a property of specific disclosure moments~\cite{nissenbaum2004privacy,petronio2002boundaries}, largely assuming that information's relevance decays as it ages out of feeds. \revise{Our data shows this assumption fails in doxxing contexts: doxxers actively mine the digital past, producing a ``temporal asymmetry'' between how survivors remember their digital history and how doxxers can retrieve it. This extends} privacy beyond moment-of-disclosure to managing persistent archives where contexts collapse across time~\cite{marwick2011tweet}, a problem compounded by the compulsory legibility (\S\ref{sec:mandatory-identity-infras}) that keeps past content anchored to the same legal person indefinitely. 

\revise{These dynamics carry concrete design implications. Beyond generic time-decay, platforms should consider: (i) \textit{graduated retrievability}, making older content harder to surface through search; and (ii) \textit{user-initiated decay}, allowing survivors to reduce indexability of historical content without proving specific harm. More broadly, privacy theory needs frameworks treating digital permanence not as a neutral default but as a design choice with distributive consequences, one that disproportionately burdens those without the resources to curate their digital past.}

Third, building on §\ref{sec:boundary-exposure} and §\ref{regulatory_mandates}, survivors experienced recognition of complete powerlessness distinct from the asymmetries documented in prior work. Literature on domestic violence~\cite{freed2018stalker} or workplace monitoring~\cite{he2025exploring} typically reveals survivors retaining residual agency (e.g., leaving abusive relationships, encrypting communications, seeking workplace protections), justifying solutions focused on user empowerment. In contrast, doxxing survivors face foreclosed agency where self-protection is impossible: users cannot refuse mandatory identification, remove disseminated data, or counter adversaries. This structural foreclosure implies that ``user empowerment'' paradigms are insufficient; protection requires constraining the systems that render self-protection impossible.


\subsection{Recommendations}
\label{RECOMMENDATION}
\textbf{Legal and Regulatory Interventions.}
Despite multiple personal information statutes (Appendix \ref{appendix-legislation}), China's legal framework lacks clear sentencing guidelines and no explicit doxxing provisions, creating enforcement ambiguity (see \S \ref{regulatory_mandates}). Hence, we recommend three legal and regulatory interventions: First, refine legal definitions to distinguish doxxing: (1) non-consensual disclosure of identifiable information (2) with intent or foreseeability of enabling harm (3) through public dissemination to an audience capable of causing harm. Second, based on findings in \S \ref{risk_social_media}, secondary participants in doxxing attacks often face no consequences, creating ``low-risk, high-reward'' conditions. 
We recommend tiered penalties based on information sensitivity, audience reach, and documented harm, extending to those who amplify or repost exposed data. Legal frameworks should recognize networked harm as collectively produced, requiring collective accountability. This would help clarify the application of Criminal Law Article 253 and reduce enforcement discretion. 
Lastly, Chinese regulations should draw 
on International frameworks (e.g., Australia's Online Safety Act~\cite{Australia} and EU Digital Services Act~\cite{EU-digitalservice}) to mandate platform responsibilities such as (1) rapid takedown of doxxing content within defined timeframes (as \cite{truong2025delayed} shows, delays undermine moderation); (2) proactive detection of obvious PII exposure, and (3) transparent case management that offering survivors status updates and outcome summarize to ensure acknowledgement and closure. 

\textbf{Platform and Technical Interventions.} \revise{Our findings suggest that platforms should intervene across the doxxing attack chain rather than treating doxxing as an isolated content violation. (1) Platforms should reduce default user legibility by minimizing unnecessary metadata visibility, including overly granular IP-location labels, follower lists, precise timestamps, profile-creation dates, and account-recovery hints, since survivors described these fragments as being chained together for identity reconstruction (§\ref{sec:aggregated-exposure}, §\ref{regulatory_mandates}). (2) Platforms should constrain cross-platform aggregation by decoupling public identifiers from real-name verification artifacts, limiting exposure of stable identifiers such as phone numbers and reused usernames, and rate-limiting APIs that enable bulk profile lookup. (3) Because secondary amplification produced much of the harm (§\ref{risk_social_media}), moderation should operate at the cascade level: adjudicated doxxing content and near-duplicate reposts or screenshots should trigger friction, demotion, removal, or graduated penalties for amplifiers as well as original posters. Such cascade-aware mechanisms operationalize prior work on amplification dynamics~\cite{milli2025engagement,huszar2022algorithmic,truong2025delayed}, while positive-reinforcement moderation may further help by elevating prosocial content and making anti-doxxing norms visible~\cite{lambert2024positive}. Platforms could also use pre-publication PII detection~\cite{karimi2022automated} and friction-based warnings~\cite{im2020synthesized}, while preserving manual review, appeals, and safeguards for counterspeech, evidence preservation, and public-interest reporting. (4) Platform-native AI features should include rate limits, purpose verification, and refusal protocols for prompts that infer third-party identifiers, given participants' concern that LLM-enabled search and summarization lower the cost of identity reconstruction (§\ref{sec:boundary-exposure}); Luo et al.~\cite{luo2025doxxing} similarly show that LLMs can infer location-related information from photos, enabling doxxing. These measures cannot eliminate doxxing, especially under real-name infrastructures and grey-market data access, but they raise the operational cost of opportunistic attacks and target multiple points in the framework identified in §\ref{sec:conceptualising-doxxing}.}

\textbf{Social Interventions.}
As Slupska and Strohmayer argued~\cite{slupska2022networks}, effective responses to technology-facilitated harm must move beyond technical fixes to integrate care, education, and community support within a ``network of care'', one that is attentive to survivors’ psychosocial experiences~\cite{darkdoxxing} and is grounded in trauma-informed approaches~\cite{chen2026harm}. Building on prior work~\cite{he2025privacy} that highlights the need to incorporate privacy education into compulsory curricula for younger age groups, we further recommend establishing dedicated hotlines and peer support networks for doxxing survivors. These services should be staffed by trained personnel capable of addressing trauma-specific needs and complemented by accessible legal aid programs.

\subsection{Limitations}

\label{sec:limitations}
We focus on survivors because they are uniquely positioned to explain how doxxing is experienced, escalates, and persists beyond the initial incident. With this scope in mind, several limitations should be noted. As with all qualitative research, our findings rely on self-reported data shaped by participants' perceptions and recollections, which may introduce social-desirability biases~\cite{nederhof1985methods}. To mitigate this, we used open-ended, neutrally phrased questions, built rapport, reassured participants about confidentiality and anonymity, and elicited concrete episodes. 
As a qualitative study, the framework presented should be interpreted as an analytic understanding grounded in survivors' lived experiences. Follow-up quantitative studies could consider the prevalence of the patterns we identify. \change{Our cross-context comparisons draw on evidence provided by prior literature rather than a direct comparative study; the conditions discussed in §5.3 may also occur elsewhere. Future comparative research could examine how these patterns vary across local infrastructures and social practices.}
Furthermore, our sample demographics were primarily aged between 18 and 45, with 
higher educational attainment, aligning with broader trends in China's digital landscape, where social media engagement is concentrated among working-age adults rather than seniors\footnote{China report estimated that by early 2025, social media engagement is heavily concentrated among working-age adults rather than seniors (\url{https://datareportal.com/reports/digital-2025-china}).}. 
\change{This sample may underrepresent the experiences of older adults and people with lower educational attainment. Older adults} 
could experience distinct forms of harm due to lower digital literacy or reduced awareness of legal recourse, so future research could explore a broader range of survivor groups and other stakeholders, such as platform moderators and law enforcement officials.

\section{Conclusion}

This paper presents one of the first empirical studies of doxxing within China's socio-technical ecosystem, revealing how platform infrastructures, real-name identity regimes, and cultural norms collectively lower the cost of exposure and amplify harm. Through interviews with 18 survivors, we show that doxxing fundamentally reshapes users' privacy and security models---turning privacy from a personal preference into a matter of safety, structural vulnerability, and long-term risk.

Our findings demonstrate that Chinese social media platforms, by design, make doxxing easy to execute and difficult to contain, while cultural framings such as moralized vigilantism and fandom mobilization normalize these practices. We outline a multi-level response, including reducing default user legibility, implementing cascade-aware moderation, clarifying legal protections, and strengthening social interventions such as privacy education. By situating doxxing within its broader socio-technical context, this work advances understanding of technology-facilitated privacy attacks and underscores the need for systemic, survivor-centered protections.

\section*{Ethical Considerations}
\label{sec:ethics}

Doxxing may already have caused potential trauma to survivors (participants involved in this study). Therefore, careful consideration must be given to the design and conduct of interviews to uphold ethical standards, as revisiting these events may evoke emotional distress or reawaken past harms~\cite{alessi2023toward,seedat2004ethics}. However, we are not the first to engage in this type of research. Previous studies~\cite{abuse2014samhsa,portman2020implementing,kirst2017provider,neria2002trauma,thomas2019trauma,chen2022trauma,center2014trauma,razi2024toward}, particularly within the field of HCI (Human-Computer Interaction) and UPS (Usable Privacy and Security) ~\cite{thomas2019trauma,chen2022trauma,razi2024toward,ramjit2024navigating}, have contributed valuable insights and practical strategies for applying trauma-informed principles to research involving sensitive topics and vulnerable populations.

Building upon these works, we incorporated the core principles of the trauma-informed framework into our research design. Specifically, we adopted the Six Key Principles proposed by the CDC\footnote{\url{https://stacks.cdc.gov/view/cdc/56843}} and the Substance Abuse and Mental Health Services Administration (SAMHSA)~\cite{center2014trauma}. \textbf{\textit{(i) Safety}} was prioritized by giving participants full control over their participation, including the ability to pause or withdraw at any time. We also avoided questions likely to cause unnecessary emotional distress and conducted interviews in secure, private settings. \textbf{\textit{(ii) Trustworthiness and Transparency}} were ensured through the use of clear, accessible consent procedures and detailed explanations of data use. We emphasized that all identifiable information would be removed or anonymized prior to analysis and publication. \textbf{\textit{(iii) Peer Support}} was acknowledged by describing doxxing as a widespread experience, helping participants feel less isolated in sharing their experiences. \textbf{\textit{(iv) Collaboration and Mutuality}} informed our approach by treating participants as experts in their own lived experiences and inviting them to shape how their narratives would be represented. \textit{\textbf{(v) Empowerment, Voice, and Choice}} were embedded by allowing participants to control the depth and direction of their disclosure and by avoiding prescriptive or invasive lines of questioning.
\textbf{\textit{(vi) Cultural, Historical, and Gender Responsiveness}} was reflected in our attention to culturally and historically situated norms 
that shape doxxing practices in Chinese social media. We also committed to using inclusive, non-pathologizing language throughout.

\revise{All participants provided informed consent before interviews began. The consent process covered the study purpose, interview topics, recording and transcription, anonymization, data storage and use, and participants’ right to pause, skip questions, or withdraw without losing compensation. Audio recordings were stored on password-protected, encrypted research devices and institutional servers accessible only to the research team. Transcripts were de-identified before analysis.}

\revise{\textbf{\textit{Handling Sensitive Disclosures of Counter-Responses}}. Because our study sought to understand the diverse ways in which survivors processed and responded to doxxing, the screener included a neutral question about whether participants had considered or attempted retaliatory doxxing. Since questions about sensitive or potentially harmful behaviours require careful design, we drew on ethics guidance emphasizing neutral wording, participant control, and avoidance of suggestive framing~\cite{sexual2012ethical,dazzi2014does}. The question therefore treated counter-doxxing as one possible response rather than an expected, justified, or effective one; It avoided revenge- or justice-oriented language, did not ask for motives, methods, or technical details, and allowed participants to skip the question. Across both screening responses and interviews, no participant reported publicly sharing their doxxer’s private information; disclosed cases involved only consideration or unsuccessful attempts.}

\revise{\textbf{\textit{Broader Stakeholder Considerations.}} We also considered other stakeholders beyond our participants, including their doxxers, individuals targeted by retaliatory attempts, and broader affected communities. The key risks were enabling doxxing through operational details, re-identifying people involved in specific incidents, or stigmatizing communities such as fandom groups. We mitigated these risks through aggregation, non-attribution, and excluding technical or procedural details. Our recommendations focus on survivor protection rather than punishment or retaliation.}

In addition to these principles, we also offered referrals and informational resources, such as mental health hotlines and legal aid services, when appropriate, following the recommendations of Muraglia et al.~\cite{muraglia2020conducting} for conducting interviews on sensitive topics. Similarly, following Bellini et al.~\cite{bellini2024sok}, when working with at-risk participants, it is important to design recruitment, consent, and interview procedures that minimize potential harm, avoid dependence on employers or gatekeeping agencies, and provide participants with clear and accessible ways to withdraw at any time. Recognizing that researchers working with traumatic or sensitive content may experience emotional fatigue or distress~\cite{mckenzie2017ethical,vidgen2019challenges}, we implemented regular emotional check-ins, fostered mutual support within the team, ensured a reasonable distribution of workload, and provided access to psychological support resources when needed. Our method and procedures were approved by the Research Ethics Committee at our institution (Ethics ID: MRSP-24/25-50867).


\section*{Acknowledgment}

We would like to thank all the participants who shared their experiences and generously offered their support and willingness to help other survivors in similar situations through this study. We are also grateful to the anonymous reviewers and our shepherd for their insightful feedback. Finally, we extend our gratitude to the members of the HASP Lab for their valuable comments on earlier drafts of this paper. This research was partially supported by Generalitat Valenciana project PROMETEO CIPROM/2023/23 and Grant PID2023-151536OB-100 funded by MICIU/AEI/10.13039/501100011033 and by ERDF/EU.



\bibliographystyle{IEEEtran}
\bibliography{sample-base}

@String{Computing = "Computing" }

@String{Computer = "{IEEE} Computer" }

@String{Springer = "Springer-Verlag" }

@inproceedings{herbert2023world,
  title={A world full of privacy and security (mis) conceptions? Findings of a representative survey in 12 countries},
  author={Herbert, Franziska and Becker, Steffen and Schaewitz, Leonie and Hielscher, Jonas and Kowalewski, Marvin and Sasse, Angela and Acar, Yasemin and D{\"u}rmuth, Markus},
  booktitle={Proceedings of the 2023 CHI Conference on Human Factors in Computing Systems},
  pages={1--23},
  year={2023}
}

@inproceedings{wang2011concerned,
  title={Who Is Concerned about What? A Study of American, Chinese and Indian Users’ Privacy Concerns on Social Network Sites: (Short Paper)},
  author={Wang, Yang and Norice, Gregory and Cranor, Lorrie Faith},
  booktitle={International conference on trust and trustworthy computing},
  pages={146--153},
  year={2011},
  organization={Springer}
}

@article{seedat2004ethics,
  title={Ethics of research on survivors of trauma},
  author={Seedat, Soraya and Pienaar, Willem P and Williams, David and Stein, Daniel J},
  journal={Current psychiatry reports},
  volume={6},
  number={4},
  pages={262--267},
  year={2004},
  publisher={Springer}
}

@article{wang2015privacy,
  title={Privacy trust crisis of personal data in China in the era of Big Data: The survey and countermeasures},
  author={Wang, Zhong and Yu, Qian},
  journal={Computer Law \& Security Review},
  volume={31},
  number={6},
  pages={782--792},
  year={2015},
  publisher={Elsevier}
}

@article{huey2025cops,
  title={“Cops Need Doxxed”: Releasing Personal Information of Police Officers as a Tool of Political Harassment},
  author={Huey, Laura and Ferguson, Lorna and Towns, Zachary},
  journal={Crime \& Delinquency},
  volume={71},
  number={3},
  pages={714--739},
  year={2025},
  publisher={SAGE Publications Sage CA: Los Angeles, CA}
}

@article{franz2023doxing,
  title={Doxing and doxees: A qualitative analysis of victim experiences and responses},
  author={Franz, Anjuli and Thatcher, Jason Bennett},
  year={2023}
}

@inproceedings{wang2016examining,
  title={Examining American and Chinese internet users' contextual privacy preferences of behavioral advertising},
  author={Wang, Yang and Xia, Huichuan and Huang, Yun},
  booktitle={Proceedings of the 19th ACM conference on computer-supported cooperative work \& social computing},
  pages={539--552},
  year={2016}
}

@inproceedings{chen2023we,
  title={How we express ourselves freely: Censorship, self-censorship, and anti-censorship on a Chinese social media},
  author={Chen, Xiang and Xie, Jiamu and Wang, Zixin and Shen, Bohui and Zhou, Zhixuan},
  booktitle={International Conference on Information},
  pages={93--108},
  year={2023},
  organization={Springer}
}

@article{zhang2023understanding,
  title={Understanding privacy over-collection in wechat sub-app ecosystem},
  author={Zhang, Xiaohan and Wang, Yang and Zhang, Xin and Huang, Ziqi and Zhang, Lei and Yang, Min},
  journal={arXiv preprint arXiv:2306.08391},
  year={2023}
}

@article{lambert2024positive,
  title={" Positive reinforcement helps breed positive behavior": Moderator Perspectives on Encouraging Desirable Behavior},
  author={Lambert, Charlotte and Choi, Frederick and Chandrasekharan, Eshwar},
  journal={Proceedings of the ACM on Human-Computer Interaction},
  volume={8},
  number={CSCW2},
  pages={1--33},
  year={2024},
  publisher={ACM New York, NY, USA}
}

@inproceedings{im2020synthesized,
  title={Synthesized social signals: Computationally-derived social signals from account histories},
  author={Im, Jane and Tandon, Sonali and Chandrasekharan, Eshwar and Denby, Taylor and Gilbert, Eric},
  booktitle={Proceedings of the 2020 CHI Conference on Human Factors in Computing Systems},
  pages={1--12},
  year={2020}
}

@article{choi2023convex,
  title={ConvEx: A Visual Conversation Exploration System for Discord Moderators},
  author={Choi, Frederick and Bajpai, Tanvi and Pratipati, Sowmya and Chandrasekharan, Eshwar},
  journal={Proceedings of the ACM on Human-Computer Interaction},
  volume={7},
  number={CSCW2},
  pages={1--30},
  year={2023},
  publisher={ACM New York, NY, USA}
}

@article{alessi2023toward,
  title={Toward a trauma-informed qualitative research approach: Guidelines for ensuring the safety and promoting the resilience of research participants},
  author={Alessi, Edward J and Kahn, Sarilee},
  journal={Qualitative Research in Psychology},
  volume={20},
  number={1},
  pages={121--154},
  year={2023},
  publisher={Taylor \& Francis}
}

@article{razi2024toward,
  title={Toward trauma-informed research practices with youth in hci: Caring for participants and research assistants when studying sensitive topics},
  author={Razi, Afsaneh and Seberger, John S and Alsoubai, Ashwaq and Naher, Nurun and De Choudhury, Munmun and Wisniewski, Pamela J},
  journal={Proceedings of the ACM on human-computer interaction},
  volume={8},
  number={CSCW1},
  pages={1--31},
  year={2024},
  publisher={ACM New York, NY, USA}
}

@article{abuse2014samhsa,
  title={SAMHSA’s concept of trauma and guidance for a trauma-informed approach},
  author={Abuse, Substance and others},
  year={2014},
  publisher={Substance Abuse and Mental Health Services Administration}
}

@article{portman2020implementing,
  title={Implementing trauma-informed care in mental health services},
  author={Portman-Thompson, Kate},
  journal={Mental Health Practice},
  volume={23},
  number={3},
  year={2020},
  publisher={RCN Publishing Company Limited}
}

@article{zhou2024ethics,
  title={Ethics of doxxing and cyberbullying in dermatology},
  author={Zhou, Albert E and Rao, Ishani H and Jain, Neelesh P and Gronbeck, Christian and Sloan, Brett and Grant-Kels, Jane M and Feng, Hao},
  journal={Clinics in dermatology},
  volume={42},
  number={6},
  year={2024},
  publisher={Elsevier}
}

@article{luo2025doxxing,
  title={doxxing via the Lens: Revealing Location-related Privacy Leakage on Multi-modal Large Reasoning Models},
  author={Luo, Weidi and Lu, Tianyu and Zhang, Qiming and Liu, Xiaogeng and Hu, Bin and Zhao, Yue and Zhao, Jieyu and Gao, Song and McDaniel, Patrick and Xiang, Zhen and others},
  journal={arXiv preprint arXiv:2504.19373},
  year={2025}
}

@book{hinduja2014bullying,
  title={Bullying beyond the schoolyard: Preventing and responding to cyberbullying},
  author={Hinduja, Sameer and Patchin, Justin W},
  year={2014},
  publisher={Corwin press}
}

@inproceedings{vidgen2019challenges,
  title={Challenges and frontiers in abusive content detection},
  author={Vidgen, Bertie and Harris, Alex and Nguyen, Dong and Tromble, Rebekah and Hale, Scott and Margetts, Helen},
  booktitle={Proceedings of the third workshop on abusive language online},
  year={2019},
  organization={Association for Computational Linguistics}
}

@article{mckenzie2017ethical,
  title={Ethical considerations in sensitive suicide research reliant on non-clinical researchers},
  author={Mckenzie, Sarah K and Li, Cissy and Jenkin, Gabrielle and Collings, Sunny},
  journal={Research ethics},
  volume={13},
  number={3-4},
  pages={173--183},
  year={2017},
  publisher={SAGE Publications Sage UK: London, England}
}

@article{center2014trauma,
  title={Trauma-informed care in behavioral health services},
  author={Center for Substance Abuse Treatment and others},
  year={2014},
  publisher={Substance Abuse and Mental Health Services Administration (US)}
}

@inproceedings{chen2022trauma,
  title={Trauma-informed computing: Towards safer technology experiences for all},
  author={Chen, Janet X and McDonald, Allison and Zou, Yixin and Tseng, Emily and Roundy, Kevin A and Tamersoy, Acar and Schaub, Florian and Ristenpart, Thomas and Dell, Nicola},
  booktitle={Proceedings of the CHI conference on human factors in computing systems},
  year={2022}
}

@article{thomas2019trauma,
  title={Trauma-informed practices in schools across two decades: An interdisciplinary review of research},
  author={Thomas, M Shelley and Crosby, Shantel and Vanderhaar, Judi},
  journal={Review of research in education},
  year={2019},
  publisher={SAGE Publications Sage CA: Los Angeles, CA}
}

@article{neria2002trauma,
  title={Trauma exposure and posttraumatic stress disorder in psychosis: findings from a first-admission cohort.},
  author={Neria, Yuval and Bromet, Evelyn J and Sievers, Sylvia and Lavelle, Janet and Fochtmann, Laura J},
  journal={Journal of consulting and Clinical Psychology},
  volume={70},
  number={1},
  pages={246},
  year={2002},
  publisher={American Psychological Association}
}

@article{kirst2017provider,
  title={Provider and consumer perceptions of trauma informed practices and services for substance use and mental health problems},
  author={Kirst, Maritt and Aery, Anjana and Matheson, Flora I and Stergiopoulos, Vicky},
  journal={International Journal of Mental Health and Addiction},
  volume={15},
  number={3},
  pages={514--528},
  year={2017},
  publisher={Springer}
}

@article{truong2025delayed,
  title={Delayed takedown of illegal content on social media makes moderation ineffective},
  author={Truong, Bao Tran and Kim, Sangyeon and Nogara, Gianluca and Verdolotti, Enrico and Sahneh, Erfan Samieyan and Saurwein, Florian and Just, Natascha and Luceri, Luca and Giordano, Silvia and Menczer, Filippo},
  journal={arXiv preprint arXiv:2502.08841},
  year={2025}
}

@article{braun2006using,
  title={Using thematic analysis in psychology},
  author={Braun, Virginia and Clarke, Victoria},
  journal={Qualitative research in psychology},
  volume={3},
  number={2},
  pages={77--101},
  year={2006},
  publisher={Taylor \& Francis}
}

@article{rasheed2024influence,
  title={Influence Operations},
  author={Rasheed, Adil},
  journal={The Sharp Power of Non-Kinetic Subversion, New Delhi},
  year={2024}
}

@article{anderson2022harm,
  title={Harm imbrication and virtualised violence: Reconceptualising the harms of doxxing},
  author={Anderson, Briony and Wood, Mark A},
  journal={International Journal for Crime, Justice and Social Democracy},
  year={2022},
  publisher={Queensland University of Technology. Crime and Justice Research Centre~…}
}

@inproceedings{ramjit2024navigating,
  title={Navigating traumatic stress reactions during computer security interventions},
  author={Ramjit, Lana and Dolci, Natalie and Rossi, Francesca and Garcia, Ryan and Ristenpart, Thomas and Cuomo, Dana},
  booktitle={33rd USENIX Security Symposium (USENIX Security 24)},
  pages={2011--2028},
  year={2024}
}

@article{eckert2020doxxing,
  title={Doxxing, privacy and gendered harassment. The shock and normalization of veillance cultures},
  author={Eckert, Stine and Metzger-Riftkin, Jade},
  journal={M\&K Medien \& Kommunikationswissenschaft},
  volume={68},
  number={3},
  year={2020},
  publisher={Nomos Verlagsgesellschaft mbH \& Co. KG}
}

@inproceedings{gupta2024really,
  title={" I really just leaned on my community for support": Barriers, Challenges, and Coping Mechanisms Used by Survivors of $\{$Technology-Facilitated$\}$ Abuse to Seek Social Support},
  author={Gupta, Naman and Walsh, Kate and Das, Sanchari and Chatterjee, Rahul},
  booktitle={33rd USENIX Security Symposium (USENIX Security 24)},
  pages={4981--4998},
  year={2024}
}

@inproceedings{slupska2022networks,
  title={Networks of care: Tech abuse advocates' digital security practices},
  author={Slupska, Julia and Strohmayer, Angelika},
  booktitle={31st USENIX Security Symposium (USENIX Security 22)},
  pages={341--358},
  year={2022}
}

@misc{ukgov,
  author        = "Cyber security breaches survey 2025",
  year          = 2025,
  title         = "Cyberbullying'",
  url           = 
"https://www.gov.uk/government/statistics/cyber-security-breaches-survey-2025/cyber-security-breaches-survey-2025#chapter-6-cyber-crime",
  month         = mar,
  lastaccessed  = "Nov 22, 2025",
}

@article{nissenbaum2004privacy,
  title={Privacy as contextual integrity},
  author={Nissenbaum, Helen},
  journal={Wash. L. Rev.},
  volume={79},
  pages={119},
  year={2004},
  publisher={HeinOnline}
}

@inproceedings{freed2018stalker,
  title={“a stalker's paradise” how intimate partner abusers exploit technology},
  author={Freed, Diana and Palmer, Jackeline and Minchala, Diana and Levy, Karen and Ristenpart, Thomas and Dell, Nicola},
  booktitle={Proceedings of the 2018 CHI conference on human factors in computing systems},
  pages={1--13},
  year={2018}
}

@article{marwick2011tweet,
  title={I tweet honestly, I tweet passionately: Twitter users, context collapse, and the imagined audience},
  author={Marwick, Alice E and Boyd, Danah},
  journal={New media \& society},
  volume={13},
  number={1},
  pages={114--133},
  year={2011},
  publisher={Sage Publications Sage UK: London, England}
}

@incollection{trepte2014people,
  title={Do people know about privacy and data protection strategies? Towards the “Online Privacy Literacy Scale”(OPLIS)},
  author={Trepte, Sabine and Teutsch, Doris and Masur, Philipp K and Eicher, Carolin and Fischer, Mona and Hennh{\"o}fer, Alisa and Lind, Fabienne},
  booktitle={Reforming European data protection law},
  pages={333--365},
  year={2014},
  publisher={Springer}
}

@inproceedings{he2025privacy,
  title={Privacy Perspectives and Practices of Chinese Smart Home Product Teams},
  author={He, Shijing and Lei, Yaxiong and Zhan, Xiao and Zhang, Chi and Ye, Juan and Abu-Salma, Ruba and Such, Jose},
  booktitle={2026 IEEE Symposium on Security and Privacy (SP)},
  year={2026},
  organization={IEEE}
}

@article{westin1968privacy,
  title={Privacy and freedom},
  author={Westin, Alan F},
  journal={Washington and Lee Law Review},
  volume={25},
  number={1},
  pages={166},
  year={1968}
}

@misc{Australia,
  author        = "",
  year          = 2021,
  title         = "Online Safety Act 2021",
  url           = 
"https://www.legislation.gov.au/C2021A00076/latest/text",
  month         = mar,
  lastaccessed  = "Nov 22, 2025",
}

@article{darkdoxxing,
title = {Dark doxxing: How Dark Triad traits impact support for doxxing behaviors},
journal = {Personality and Individual Differences},
volume = {217},
pages = {112432},
year = {2024},
issn = {0191-8869},
doi = {https://doi.org/10.1016/j.paid.2023.112432},
url = {https://www.sciencedirect.com/science/article/pii/S0191886923003550},
author = {Stephen Foster and Jasmine Cross}
}

@article{braun2019reflecting,
  title={Reflecting on reflexive thematic analysis},
  author={Braun, Virginia and Clarke, Victoria},
  journal={Qualitative research in sport, exercise and health},
  year={2019},
  publisher={Taylor \& Francis}
}

@misc{EU-digitalservice,
  author        = "",
  year          = 2021,
  title         = "The Digital Services Act (DSA)",
  url           = 
"https://eur-lex.europa.eu/legal-content/EN/TXT/?uri=celex%3A32022R2065",
  month         = mar,
  lastaccessed  = "Nov 22, 2025",
}

@article{kim2022does,
  title={When does it become harassment? An investigation of online criticism and calling out in Twitter},
  author={Kim, Haesoo and Kim, HaeEun and Kim, Juho and Jang, Jeong-woo},
  journal={Proceedings of the ACM on Human-Computer Interaction},
  volume={6},
  number={CSCW2},
  pages={1--32},
  year={2022},
  publisher={ACM New York, NY, USA}
}

@article{kim2024respect,
  title={ReSPect: Enabling Active and Scalable Responses to Networked Online Harassment},
  author={Kim, Haesoo and Lee, Juhoon and Jang, Jeong-Woo and Kim, Juho},
  journal={Proceedings of the ACM on Human-Computer Interaction},
  year={2024},
  publisher={ACM New York, NY, USA}
}

@article{huszar2022algorithmic,
  title={Algorithmic amplification of politics on Twitter},
  author={Husz{\'a}r, Ferenc and Ktena, Sofia Ira and O’Brien, Conor and Belli, Luca and Schlaikjer, Andrew and Hardt, Moritz},
  journal={Proceedings of the national academy of sciences},
  year={2022},
  publisher={National Academy of Sciences}
}

@inproceedings{thomas2022s,
  title={“It’s common and a part of being a content creator”: Understanding How Creators Experience and Cope with Hate and Harassment Online},
  author={Thomas, Kurt and Kelley, Patrick Gage and Consolvo, Sunny and Samermit, Patrawat and Bursztein, Elie},
  booktitle={Proceedings of the 2022 CHI conference on human factors in computing systems},
  pages={1--15},
  year={2022}
}

@inproceedings{kumar2023understanding,
  title={Understanding the behaviors of toxic accounts on reddit},
  author={Kumar, Deepak and Hancock, Jeff and Thomas, Kurt and Durumeric, Zakir},
  booktitle={Proceedings of the ACM web conference 2023},
  pages={2797--2807},
  year={2023}
}

@article{milli2025engagement,
  title={Engagement, user satisfaction, and the amplification of divisive content on social media},
  author={Milli, Smitha and Carroll, Micah and Wang, Yike and Pandey, Sashrika and Zhao, Sebastian and Dragan, Anca D},
  journal={PNAS nexus},
  year={2025},
  publisher={Oxford University Press US}
}

@book{petronio2002boundaries,
  title={Boundaries of privacy: Dialectics of disclosure},
  author={Petronio, Sandra},
  year={2002},
  publisher={Suny Press}
}

@misc{amada-def,
  author        = "Amanda B",
  year          = 2012,
  title         = "doxxing Meme'",
  day           = 5,
  url           = 
"https://knowyourmeme.com/memes/doxxing",
  month         = mar,
  lastaccessed  = "July 1, 2025",
}

@misc{death-threat,
  author        = "Jasmine McNealy",
  year          = 2018,
  title         = "What is doxxing, and why is it so scary?",
  day           = 5,
  url           = 
"https://theconversation.com/what-is-doxxing-andwhy-is-it-so-scary-95848",
  month         = mar,
  lastaccessed  = "July 1, 2025",
}

@article{dazzi2014does,
  title={Does asking about suicide and related behaviours induce suicidal ideation? What is the evidence?},
  author={Dazzi, Tommaso and Gribble, Rachael and Wessely, Simon and Fear, Nicola T},
  journal={Psychological medicine},
  volume={44},
  number={16},
  pages={3361--3363},
  year={2014},
  publisher={Cambridge University Press}
}

@book{sexual2012ethical,
  title={Ethical and safety recommendations for research on perpetration of sexual violence},
  author={Sexual Violence Research Initiative and Jewkes, Rachel and Dartnall, Elizabeth and Sikweyiya, Yandisa and others},
  year={2012},
  publisher={Sexual Violence Research Initiative, Medical Research Council Pretoria~…}
}

@incollection{stanfill2024fandom,
  title={Fandom is ugly: Networked harassment in participatory culture},
  author={Stanfill, Mel},
  booktitle={Fandom Is Ugly},
  year={2024},
  publisher={New York University Press}
}

@article{marwick2021morally,
  title={Morally motivated networked harassment as normative reinforcement},
  author={Marwick, Alice E},
  journal={Social Media+ Society},
  volume={7},
  number={2},
  pages={20563051211021378},
  year={2021},
  publisher={SAGE Publications Sage UK: London, England}
}

@article{ong2025hong,
  title={Hong Kong's response to the fight against doxxing},
  author={Ong, Rebecca},
  journal={Common Law World Review},
  volume={54},
  number={1},
  pages={17--42},
  year={2025},
  publisher={Sage Publications Sage UK: London, England}
}

@misc{beaujon-def,
  author        = "Beaujon A",
  year          = 2014,
  title         = "Redditors furious Newsweek ‘doxxed’ Bitcoin Founder'",
  day           = 5,
  url           = 
"http://www.poynter.org/news/mediawire/242348/redditors-furious-newsweek-outed-bitcoin-founder/",
  month         = mar,
  lastaccessed  = "July 1, 2025",
}

@inproceedings{snyder2017fifteen,
  title={Fifteen minutes of unwanted fame: Detecting and characterizing doxxing},
  author={Snyder, Peter and Doerfler, Periwinkle and Kanich, Chris and McCoy, Damon},
  booktitle={Proceedings of the 2017 internet measurement conference},
  pages={432--444},
  year={2017}
}

@article{douglas2016doxxing,
  title={doxxing: A conceptual analysis},
  author={Douglas, David M},
  journal={Ethics and information technology},
  volume={18},
  number={3},
  pages={199--210},
  year={2016},
  publisher={Springer}
}

@article{chen2019doxxing,
  title={doxxing: What adolescents look for and their intentions},
  author={Chen, Mengtong and Cheung, Anne Shann Yue and Chan, Ko Ling},
  journal={International journal of environmental research and public health},
  volume={16},
  number={2},
  year={2019},
  publisher={MDPI}
}

@article{muraglia2020conducting,
  title={Conducting research interviews on sensitive topics},
  author={Muraglia, S and Vasquez, A and Reichert, J},
  journal={Illinois Criminal Justice Information Authority (ICJIA)},
  year={2020}
}

@article{zhou2024understanding,
  title={Understanding Chinese Internet Users' Perceptions of, and Online Platforms' Compliance with, the Personal Information Protection Law (PIPL)},
  author={Zhou, Morgana Mo and Qu, Zhiyan and Wan, Jinhan and Wen, Bo and Yao, Yaxing and Lu, Zhicong},
  journal={Proceedings of the ACM on Human-Computer Interaction},
  volume={8},
  number={CSCW1},
  pages={1--26},
  year={2024},
  publisher={ACM New York, NY, USA}
}

@article{karimi2022automated,
  title={Automated detection of doxxing on twitter},
  author={Karimi, Younes and Squicciarini, Anna and Wilson, Shomir},
  journal={Proceedings of the ACM on Human-Computer Interaction},
  volume={6},
  number={CSCW2},
  pages={1--24},
  year={2022},
  publisher={ACM New York, NY, USA}
}

@article{guest2006many,
  title={How many interviews are enough? An experiment with data saturation and variability},
  author={Guest, Greg and Bunce, Arwen and Johnson, Laura},
  journal={Field methods},
  volume={18},
  number={1},
  pages={59--82},
  year={2006},
}

@article{xian2008lost,
  title={Lost in translation? Language, culture and the roles of translator in cross-cultural management research},
  author={Xian, Huiping},
  journal={Qualitative Research in Organizations and Management: An International Journal},
  year={2008},
  publisher={Emerald Group Publishing Limited}
}

@article{braun2021saturate,
  title={To saturate or not to saturate? {Q}uestioning data saturation as a useful concept for thematic analysis and sample-size rationales},
  author={Braun, Virginia and Clarke, Victoria},
  journal={Qualitative Research in Sport, Exercise and Health},
  year={2021},
  publisher={Taylor \& Francis}
}

@article{vasileiou2018characterising,
  title={Characterising and justifying sample size sufficiency in interview-based studies: {S}ystematic analysis of qualitative health research over a 15-year period},
  author={Vasileiou, Konstantina and Barnett, Julie and Thorpe, Susan and Young, Terry},
  journal={BMC Medical Research Methodology},
  year={2018},
  publisher={Springer}
}

@inproceedings{bellini2024sok,
  title={Sok: Safer digital-safety research involving at-risk users},
  author={Bellini, Rosanna and Tseng, Emily and Warford, Noel and Daffalla, Alaa and Matthews, Tara and Consolvo, Sunny and Woelfer, Jill Palzkill and Kelley, Patrick Gage and Mazurek, Michelle L and Cuomo, Dana and others},
  booktitle={2024 IEEE Symposium on Security and Privacy (SP)},
  year={2024},
  organization={IEEE}
}

@article{chan2024confucian,
  title={How Confucian Values Shape the Moral Boundaries of Family Caregiving},
  author={Chan, HYL and Kim, Richard and Leung, DYP and Cheng, HY},
  journal={Nursing Ethics: Normative Foundations, Advanced Concepts, and Emerging Issues},
  year={2024},
  publisher={Oxford University Press}
}

@inproceedings{he2025exploring,
  title={Exploring the privacy and security challenges faced by migrant domestic workers in chinese smart homes},
  author={He, Shijing and Zhan, Xiao and Lei, Yaxiong and Liu, Yueyan and Abu-Salma, Ruba and Such, Jose},
  booktitle={Proceedings of the CHI Conference on Human Factors in Computing Systems},
  pages={1--18},
  year={2025}
}

@article{zhai2023fans,
  title={Fans’ practice of reporting: A study of the structure of data fan labor on Chinese social media},
  author={Zhai, Haoyang and Wang, Wilfred Yang},
  journal={International journal of communication},
  volume={17},
  pages={22},
  year={2023}
}

@article{dan2023observations,
  title={Observations of Chinese fandom: organizational characteristics and the relationships inside and outside the “Fan circle”},
  author={Dan, Mao and Jingya, Wang and Jiajun, Chen},
  journal={The Journal of Chinese Sociology},
  year={2023},
  publisher={Springer}
}

@article{bedford2003guilt,
  title={Guilt and shame in Chinese culture: A cross-cultural framework from the perspective of morality and identity},
  author={Bedford, Olwen and Hwang, Kwang-Kuo},
  journal={Journal for the Theory of Social Behaviour},
  year={2003},
  publisher={Wiley Online Library}
}

@article{li2004organisation,
  title={The organisation of Chinese shame concepts?},
  author={Li, Jin and Wang, Lianqin and Fischer, Kurt},
  journal={Cognition and emotion},
  volume={18},
  number={6},
  pages={767--797},
  year={2004},
  publisher={Taylor \& Francis}
}

@article{calzada2022citizens,
  title={Citizens’ data privacy in China: The state of the art of the Personal Information Protection Law (PIPL)},
  author={Calzada, Igor},
  journal={Smart Cities},
  volume={5},
  number={3},
  pages={1129--1150},
  year={2022},
  publisher={MDPI}
}

@article{creemers2022china,
  title={China’s emerging data protection framework},
  author={Creemers, Rogier},
  journal={Journal of Cybersecurity},
  volume={8},
  number={1},
  pages={tyac011},
  year={2022},
  publisher={Oxford University Press}
}

@misc{jiang2018chinese,
  title={Chinese social media and big data: big data, big brother, big profit?},
  author={Jiang, Min and Fu, King-Wa},
  journal={Policy \& Internet},
  volume={10},
  number={4},
  pages={372--392},
  year={2018},
  publisher={Wiley Online Library}
}

@inproceedings{ortloff2023different,
  title={Different researchers, different results? analyzing the influence of researcher experience and data type during qualitative analysis of an interview and survey study on security advice},
  author={Ortloff, Anna-Marie and Fassl, Matthias and Ponticello, Alexander and Martius, Florin and Mertens, Anne and Krombholz, Katharina and Smith, Matthew},
  booktitle={Proceedings of the 2023 CHI Conference on Human Factors in Computing Systems},
  pages={1--21},
  year={2023}
}

@inproceedings{chen2026harm,
  title={From Harm to Healing: Understanding Individual Resilience after Cybercrimes},
  author={Chen, Xiaowei and Tran, Mindy and Deng, Yue and Acharya, Bhupendra and Zou, Yixin},
  booktitle={Proceedings of the 2026 CHI Conference on Human Factors in Computing Systems},
  pages={1--22},
  year={2026}
}

@article{mcdonald2019reliability,
  title={Reliability and inter-rater reliability in qualitative research: Norms and guidelines for CSCW and HCI practice},
  author={McDonald, Nora and Schoenebeck, Sarita and Forte, Andrea},
  journal={Proceedings of the ACM on human-computer interaction},
  year={2019},
  publisher={ACM New York, NY, USA}
}

@article{nederhof1985methods,
  title={Methods of coping with social desirability bias: A review},
  author={Nederhof, Anton J},
  journal={European journal of social psychology},
  volume={15},
  number={3},
  pages={263--280},
  year={1985},
  publisher={Wiley Online Library}
}

%




\section{Appendix}

\begin{table*}[!t]
\centering
\fontsize{8pt}{8pt}\selectfont
\caption{Legal provisions potentially applicable to doxxing-related conduct in the People’s Republic of China (PRC)}
\label{tab:legal_doxxers}
\begin{tabular}{l|l|p{9cm}}
\toprule
\textbf{Law/Regulation} & \textbf{Articles} & \textbf{Application to doxxing} \\
\midrule
Civil Code& Art. 1032–1039 & Privacy/personal information tort: non-consensual disclosure of address, ID, photos, or private life; victims may seek injunctive relief (takedown/deletion), cessation, apology, and damages.\\
Cybersecurity Law& Art. 44, 67 & Administrative basis for penalizing illegal acquisition/sale/provision of personal information; can also target accounts/groups used to organize unlawful online activities.\\
Personal Information Protection Law & Art. 10, 25, 27–29, 44, 66 & Prohibits illegal collection/provision/disclosure of others’ personal information; limits ``reasonable scope'' use of publicly disclosed data; heightened protection for sensitive data (e.g., location, ID, address), with penalties for violations. \\
Data Security Law& Art. 32, 51, 52 & Targets illegal acquisition of data (e.g., via black markets/databases) and supports civil liability/penalties when unlawful data acquisition or handling harms individuals. \\
Public Security Administration Law & Art. 42 & Public-order sanctions for harassment, insult/defamation, stalking-like harassment, or repeated abusive/threatening communications often accompanying doxxing. \\
Criminal Law & Art. 253 (1), 246, 293 & Applies to serious illegal collection or disclosure of personal information, including organized doxxing; aggravated harassment may also trigger Arts. 246 or 293. \\
\bottomrule
\end{tabular}
\begin{minipage}{0.95\linewidth}
\footnotesize
\textit{Note.} Official texts of the listed laws are available here: Civil Code of the People's Republic of China (\hyperlink{https://www.gov.cn/xinwen/2020-06/01/content_5516649.htm}{Link}), Cybersecurity Law (\hyperlink{https://www.gov.cn/xinwen/2016-11/07/content_5129723.htm}{Link}), Personal Information Protection Law (PIPL) (\hyperlink{https://www.gov.cn/xinwen/2021-08/20/content_5632486.htm}{Link}), Data Security Law (\hyperlink{https://www.gov.cn/xinwen/2021-06/11/content_5616919.htm}{Link}), Public Security Administration Law (\hyperlink{https://www.gjxfj.gov.cn/gjxfj/xxgk/fgwj/flfg/webinfo/2016/03/1460585589901723.htm}{Link}), Criminal Law (\hyperlink{https://www.spp.gov.cn/spp/fl/201802/t20180206_364975.shtml}{Link}).
\end{minipage}
\end{table*}

\subsection{Relevant Legal Provisions}
\label{appendix-legislation}


\begin{table*}[t]
\centering
\fontsize{8pt}{8pt}\selectfont
\caption{Legal responsibilities and obligations of platforms in doxxing-related contexts in the People’s Republic of China (PRC)}
\label{tab:legal_platforms}
\begin{tabular}{l|l|p{9cm}}
\toprule
\textbf{Law/Regulation} & \textbf{Articles} & \textbf{Application to doxxing} \\
\midrule
Cybersecurity Law & Art. 41–42 & Imposes general obligations on platforms to protect personal information and to take remedial measures (e.g., deletion, blocking, notification) when personal data is unlawfully disclosed. \\
Personal Information Protection Law & Art. 55–58 & Requires platforms to implement technical and organizational measures to prevent unlawful disclosure or algorithmic amplification of users’ personal information. \\
Internet Information Service Regulations & Art. 15, 16 & Obliges platforms to restrict the upload or dissemination of unlawful content, including posts disclosing identifiable personal information. \\
Online Content Ecosystem Governance Provisions & Art. 6, 13 & Requires platforms to monitor, review, and remove unlawful or harmful content, including doxxing-related disclosures of personal information. \\
Law of the P.R.C. on the Protection of Minors & Art. 74 & Mandates prompt removal of content disclosing minors’ personal information (e.g., school, identity, family details), with heightened platform obligations. \\
\bottomrule
\end{tabular}
\begin{minipage}{0.95\linewidth}
\footnotesize
\textit{Note.} Official texts of the listed laws are available here: Internet Information Service Regulations (\hyperlink{https://www.gov.cn/gongbao/content/2000/content_60531.htm}{Link}), Online Content Ecosystem Governance Provisions (\hyperlink{https://www.cac.gov.cn/2019-12/20/c_1578375159509309.htm}{Link}), Law of the P.R.C. on the Protection of Minors (\hyperlink{https://www.gov.cn/xinwen/2020-10/18/content_5552113.htm}{Link}).
\end{minipage}
\end{table*}

This appendix provides a structured overview of legal provisions referenced in \S\ref{sec:related-mitigation} that may be applied to doxxing-related conduct in Mainland China. In the absence of a standalone anti-doxxing law, these provisions form a fragmented legal framework through which doxxing is regulated in practice. All legal texts referenced here were accessed and analyzed based on their effective versions as of November 2025. The analysis is organized into two tables corresponding to the primary actors involved in doxxing incidents: individual doxxers and online platforms hosting user-generated content.

Table \ref{tab:legal_doxxers} outlines the key articles from major Chinese laws (including the Civil Code, Cybersecurity Law, and Criminal Law) that can be applied to hold doxxers legally accountable. The table highlights how general provisions regarding personal information protection, privacy, and public security administration are utilized in practice to regulate doxxing acts.

Table \ref{tab:legal_platforms} summarizes the principal articles defining the responsibilities and obligations of online platforms in managing and responding to doxxing content. These provisions mandate actions such as content removal, prevention of information leakage, and timely response, especially in the context of protecting minors.

\subsection{Demographics}

Table \ref{tab:participant-details} presents detailed information about our interview participants.

\begin{table*}[t!]
\centering
\caption{Demographic characteristics of interview participants.}
\label{tab:participant-details}
\resizebox{1\textwidth}{!}{%
\begin{tabular}{lllllll}
\toprule
\textbf{PID} & \textbf{Age }& \textbf{Gender} & \textbf{Education Level} &\textbf{ Occupation} & \textbf{Province} & \textbf{Platform(s)} \\
\midrule

P1  & 25–34 & Male   &  Master’s  & Student            & Xinjiang       & \textbf{QQ Zone}, Bilibili, RED, Weibo, WeChat \\
P2  & 35–44 & Male   & Master’s  & Information Technology  & Shanghai  & \textbf{Dianping}, Douyin, RED, WeChat \\
P3  & 25–34 & Female & Bachelor’s  & Freelancer   & Zhejiang       & \textbf{Weibo}, Douyin, QQ, RED, WeChat \\
P4  & 18–24 & Male   & Bachelor’s  & Student    & Shandong       & \textbf{Zhiboba}, Bilibili, Douyin, Hupu, RED, Weibo, WeChat \\
P5  & 18–24 & Male   & Master’s   & Student   & Hunan          & \textbf{Tieba}, QQ, RED, WeChat, Zhihu \\
P6  & 35–44 & Male   & Bachelor’s   & Energy Sector    & Shanxi         & \textbf{Douyin}, Douban, RED, Weibo, WeChat, Zhihu \\
P7  & 18–24 & Male   & Associate Degree  & Student            & Shandong       & \textbf{WeChat}, Bilibili, Douyin, QQ \\
P8  & 25–34 & Female & Master’s   & Education Sector & Chongqing      & \textbf{Tieba, Weibo}, QQ, RED, WeChat \\
P9  & 18–24 & Female & Incomplete University & Student     & Guangdong      & \textbf{QQ Zone}, Douyin, QQ, RED, Weibo, WeChat \\
P10 & 18–24 & Female & Bachelor’s   & Education Sector& Anhui          & \textbf{Taobao}, Douyin, Kuaishou, RED, Weibo, WeChat \\
P11 & 25–34 & Female & Doctoral Degree  & Student            & Hunan          & \textbf{Tieba}, Douban, RED, Weibo, WeChat \\
P12 & 25–34 & Female & Master’s   & Student            & Anhui          & \textbf{RED, QQ Zone, WeChat}, Bilibili, Douban, Zhihu \\
P13 & 18–24 & Male   & High School & Finance Sector  & Sichuan               & \textbf{Wanba App}, Douyin, Kuaishou, RED \\
P14 & 25–34 & Male   & Bachelor’s   & Student            & Beijing        & \textbf{Kuaishou, QQ Zone}, Bilibili, Douyin, RED, WeChat \\
P15 & 25–34 & Female & Master’s   & Design\& Creative Industry & Liaoning & \textbf{WeChat}, Douyin, Kuaishou, RED \\
P16 & 18–24 & Female & Bachelor’s   & Student & Beijing & \textbf{WeChat}, Bilibili, Douyin, QQ, RED, Weibo \\
P17 & 18–24 & Female & Associate Degree   & Gaming Industry  & Fujian & \textbf{WeChat}, Bilibili, Douyin, RED, Weibo \\
P18 & 25-34 & Male & Bachelor’s   & Media\&Communication Sector  & Zhejiang & \textbf{Weibo}, Douyin, Wechat \\
\bottomrule
\multicolumn{7}{l}{\parbox{\textwidth}{\small \textbf{Note:} Listed platforms are social media platforms that participants frequently use. \par Platforms in bold indicate where participants reported experiencing or being targeted by doxxing incidents.}}
\end{tabular}}
\end{table*}


\end{document}